\documentclass[12pt]{article}
\usepackage{amsmath}
\usepackage{graphicx}
\usepackage{enumerate}
\usepackage{natbib}
\usepackage{url} 
\usepackage{authblk}
\usepackage{setspace}

\usepackage{xr}

\usepackage{geometry}
\usepackage{amsmath}
\allowdisplaybreaks[4]
\usepackage{float}
\usepackage{mathrsfs}
\usepackage{amsfonts}   
\usepackage{xcolor}
\usepackage{times}
\usepackage{bm}
\usepackage{colonequals}
\usepackage{graphicx}
\usepackage{caption}
\usepackage{enumerate}
\usepackage{cmll}
\usepackage{amsthm}
\usepackage{subfigure}
\usepackage{threeparttable}
\usepackage{booktabs}
\usepackage{multirow}
\usepackage{makecell}
\usepackage{tikz}
\usetikzlibrary{positioning,arrows.meta,quotes}
\usetikzlibrary{shapes,snakes}
\usetikzlibrary{bayesnet}
\tikzset{>=latex}
\tikzstyle{plate caption} = [caption, node distance=0, inner sep=0pt,
below left=5pt and 0pt of #1.south]
\usepackage{authblk}
\usepackage{algorithm}
\usepackage{algpseudocode}

\makeatletter
\newenvironment{breakablealgorithm}
{
		\begin{center}
			\refstepcounter{algorithm}
			\hrule height.8pt depth0pt \kern2pt
			\renewcommand{\caption}[2][\relax]{
				{\raggedright\textbf{\ALG@name~\thealgorithm} ##2\par}%
				\ifx\relax##1\relax 
				\addcontentsline{loa}{algorithm}{\protect\numberline{\thealgorithm}##2}%
				\else 
				\addcontentsline{loa}{algorithm}{\protect\numberline{\thealgorithm}##1}%
				\fi
				\kern2pt\hrule\kern2pt
			}
		}{
		\kern2pt\hrule\relax
	\end{center}
}
\makeatletter

\newtheorem{theorem}{\bf{Theorem}}

\newtheorem{remark}{\bf{Remark}}
\newtheorem{lemma}{\bf{Lemma}}

\newtheorem{example}{\bf{Example}}
\newtheorem{proposition}{\bf{Proposition}}
\newtheorem{condition}{\bf{Condition}}

\newtheorem{assumption}{\bf{Assumption}}

\newcommand{\bbR}{\mathbb{R}}

\newcommand{\var}{{\rm var}}
\newcommand{\cE}{\mathcal{E}}

\newcommand{\cV}{\mathcal{V}}

\newcommand{\cX}{\mathcal{X}}
\newcommand{\cY}{\mathcal{Y}}

\newcommand{\cC}{\mathcal{C}}

\newcommand{\cI}{\mathcal{I}}

\newcommand{\cA}{\mathcal{A}}
\newcommand{\cT}{\mathcal{T}}

\newcommand{\barbeta}{\bar{\beta}}

\newcommand{\hbeta}{\hat{\beta}}
\makeatother
\graphicspath{{./fig/}}

\def\T{{ \mathrm{\scriptscriptstyle T} }}

\usepackage{easyReview}

\usepackage{lipsum}

\newcommand\blfootnote[1]{%
	\begingroup
	\renewcommand\thefootnote{}\footnote{#1}%
	\addtocounter{footnote}{-1}%
	\endgroup
}

\begin{document}
	\title{Causal Effect Identification and Inference with Endogenous Exposures and a Light-tailed Error}
    \author[a]{Ruoyu Wang}
    \author[b$*$]{Wang Miao}

   \affil[a]{Department of Biostatistics, Harvard T.H. Chan School of Public Health}
   \affil[b]{Department of Probability and Statistics, Peking University}
   \date{}
   \maketitle
   \blfootnote{$*$ Corresponding author. Email: mwfy@pku.edu.cn}
   \begin{abstract}
   	Endogeneity poses significant challenges in causal inference across various research domains. This paper proposes a novel approach to identify and estimate causal effects in the presence of endogeneity. We consider a structural equation with endogenous exposures and an additive error term. Assuming the light-tailedness of the error term, we show that the causal effect can be identified by contrasting extreme conditional quantiles of the outcome given the exposures. Unlike many existing results, our identification approach does not rely on additional parametric assumptions or auxiliary variables. Building on the identification result, we develop a new method that estimates the causal effect using extreme quantile regression. We establish the consistency of the proposed extreme-based estimator under a general additive structural equation and demonstrate its asymptotic normality in the linear model setting. 
   	Simulations and data analysis of an automobile sale dataset show the effectiveness of our method in handling endogeneity.
   \end{abstract}
	\noindent%
	\vfill
	
\setstretch{1.7}
\section{Introduction} 
Endogeneity is common in economics, epidemiology, and medical sciences, and it refers to the phenomenon that the exposure of interest is correlated with the error term in the structural equation. It arises from various sources, including unmeasured confounders, selection bias, and measurement errors. Endogeneity significantly complicates and possibly invalidates the identification and estimation of causal effects.  While additional parametric models, such as the linear factor model, offer some solutions for dealing with endogeneity \citep{wang2017confounder,guo2022doubly,ouyang2023high,tang2023synthetic,zhou2024promises}, they often come with restrictive assumptions about the data-generating process. There exists extensive literature on addressing endogeneity utilizing auxiliary variables such as instrumental variables (IVs) \citep{lewbel1998semiparametric, newey2003instrumental,lewbel2007endogenous, small2017instrumental,  wang2018bounded} and negative controls/confounder proxies  \citep{miao2018identifying, shi2020multiply, cui2023semiparametric, tchetgen2024introduction,dukes2025using}. 
However, it is challenging to identify auxiliary variables or justify their validity in practice, and the use of invalid auxiliary variables can introduce bias in the analysis. This scenario underscores the crucial need for the development of innovative identification and inference strategies to address issues of endogeneity.

This paper proposes a novel strategy for identifying causal effects in the presence of endogeneity under an additive structural equation. 
Instead of invoking additional parametric model assumptions or auxiliary variables, 
our identifying strategy rests on the light-tailedness of the error term,
which is met with many familiar distributions including the normal distribution.
Note that the challenge for identification arises from the dependence between the error term and the exposure.
Our key observation is that the extreme quantiles of the error term are approximately independent of the exposure if the error term is light-tailed and a certain regularity condition is satisfied.
Section \ref{sec: identification} presents the formal statement of these conditions and instances where they hold.
Based on this result, we show that the causal effect can be identified by leveraging extreme conditional quantiles of the outcomes without invoking additional parametric models, auxiliary variables,  or other commonly-used assumptions such as completeness \citep{newey2003instrumental} and sparsity \citep{wang2017confounder}. 
Besides, the identification strategy admits multi-dimensional exposure. Technically, our approach directly identifies the causal effect without explicitly identifying the entire structural function. This differs from most existing identification strategies, such as the exogeneity or IV-based methods, which first identify the entire structural function and then identify the causal effect by contrasting the structural function at different exposure levels.

Our identification result motivates an extreme-based method, which estimates the causal effect in the presence of endogeneity using extreme quantile regression and can be calculated by routine quantile regression packages.
We establish a non-asymptotic error bound for the proposed extreme-based estimator, demonstrating its consistency under mild conditions. For the linear model, we also establish the asymptotic normality of the proposed extreme-based estimator.
The convergence rate of the proposed estimator may not reach the parametric rate $1/\sqrt{n}$ and is generally unknown, with $n$ being the sample size. Despite this, we show that a bootstrap approach can be employed to construct a valid confidence interval for the causal effect.  The proposed extreme-based method provides a novel inference strategy for causal effects under endogeneity. Additionally, our theoretical analysis contributes to the literature on quantile regression by revealing that extreme quantile regression is invulnerable to endogeneity when the error term is light-tailed,
which has not been previously appreciated to our knowledge. The proposed extreme-based estimator is also useful in solving problems beyond causal effect estimation in the additive structural equation.
In Supplementary Material Section \ref{app: inference invalid auxiliary}, we demonstrate how to apply the estimator to select valid auxiliary variables (IVs or negative controls) in causal inference problems.
In Sections \ref{sec: sim} and \ref{sec: automobile},  we use simulation studies and an application to an automobile sale dataset to illustrate the usefulness of our method in handling endogeneity.

\section{Identification with a Light-Tailed Error}\label{sec: identification}
Suppose  we are interested in the causal effect of a $d$-dimensional exposure $X$ on an outcome $Y$,
and the causal relationship is  characterized by the following additive structural equation,
\begin{equation}\label{eq: additive model}
	Y = f_{0}(X) + \epsilon,
\end{equation}
where $f_{0}$ is the average structural function that captures the causal influence of the exposure $X$ on the outcome $Y$ and $\epsilon$ is an additive error term. 
Throughout the paper, we assume that $\epsilon$ is a mean-zero continuous random variable with a strictly increasing distribution function. 
Each component of $X$ can be either discrete or continuous. 
Model \eqref{eq: additive model} is generic and allows the function form of $f_{0}$ to be fully unspecified. It is commonly adopted in the context of statistical inference with endogenous exposures \citep{newey2003instrumental, carneiro2009estimating} and includes the widely used linear and partially linear structural equation as special cases \citep{anderson1949estimation, rothenhausler2018causal, schultheiss2023ancestor}.

Let $\cX,\cE,\cY$ be the support of $X,\epsilon$ and $Y$, respectively.
For any random variables $G_{1}$ and $G_{2}$, let $p_{G_{1}}(g_{1})$ be the density of $G_{1}$ with respect to some dominance measure and $p_{G_{1}\mid G_{2}}(g_{1}\mid g_{2})$ be the density of $G_{1}$ conditional on $G_{2} = g_{2}$. Throughout the paper, let $c$ and $C$ be generic positive constants whose values can change from place to place. For any two positive sequences $a_{n}$ and $b_{n}$, we denote $a_{n} \asymp b_{n}$ if $c \leq b_{n} / a_{n} \leq C$ for some constants $c$ and $C$.
For any $x, x_{0}$ in the support $\cX$ of $X$, let $\theta(x, x_{0}) = f_{0}(x) - f_{0}(x_{0})$ be the causal effect of the exposure level $x$ compared to the reference level $x_{0}$. 

In practice, there might be exogenous or endogenous covariates in addition to the exposure of interest. Covariates can be straightforwardly included in the vector $X$, and all results presented in this paper still hold. However, the parameter of interest may vary in the presence of covariates. For clarity, we focus on scenarios where $X$ comprises endogenous exposures. The case involving covariates is discussed in Supplementary Material Section \ref{app: adjust for covariates}.

For identification of the causal effect, the exogeneity assumption is widely adopted in empirical studies, i.e,  $X\Perp\epsilon$,  
which implies that $f_{0}(X) = E(Y\mid X)$ and further identify the causal effect by $\theta(x, x_{0})=E(Y\mid X = x) - E(Y\mid X = x_{0})$. 
However, in many real-world applications, 
unmeasured confounders, selection bias, or measurement errors arise, which render the exposure endogenous, i.e., the exposure $X$ is correlated with the error term $\epsilon$. 
For instance, in epidemiological and genetic studies, both the exposure and error term may be influenced by unmeasured factors such as population stratification or environmental and lifestyle variables, leading the exposure to be endogenous.
In the presence of endogeneity, the conditional mean $E(Y\mid X = x)$ is biased from  $f_{0}(x)$.
Therefore, identification under the exogeneity assumption 
is no longer valid and it is crucial to adjust for endogeneity.


In this paper, we consider the identification of the causal effect under endogeneity leveraging extreme outcomes.
Instead of invoking exogeneity, additional parametric models, or auxiliary variables, our strategy rests on the following two identification assumptions.

\begin{assumption}[Light-tailedness]\label{ass: tail prob}
	For any $\Delta > 0$, we have $P(\epsilon > t ) = o(P(\epsilon > t - \Delta ))$ as $t \to \infty$.
\end{assumption}
\begin{assumption}\label{ass: balance}
	For any $x\in \cX$, there are some constants $0 < c_{x} < C_{x}$ such that $c_{x} \leq p_{\epsilon\mid X}(e\mid x) / p_{\epsilon}(e) \leq C_{x}$ for any $e \in \cE$.
\end{assumption}

Assumption \ref{ass: tail prob} concerns about the upper tail probability of the error term $\epsilon$. It assume the distribution function of $\epsilon$ is short-tailed in the sense of \cite{rojo1996tail}.
Assumption \ref{ass: tail prob} holds if the tail probability of $\epsilon$ decays fast enough, in particular, when $\epsilon$ follows a bounded distribution such as uniform distribution or beta distribution. 
Assumption \ref{ass: tail prob} can be satisfied by many common unbounded light-tailed distributions such as the normal distribution, which is extensively used in medical, epidemiological, and genetic research. For an unbounded  $\epsilon$ with the decay rate  $P(\epsilon > t) \asymp \exp(- t^{a})$, 
Assumption \ref{ass: tail prob} is met provided $a > 1$. For example, Assumption \ref{ass: tail prob} is satisfied by the Rayleigh distribution and, more generally, any Weibull distribution with a shape parameter larger than one. The Weibull distribution is common in applications relevant to survival times. On the other hand, note that $P(\epsilon > t) = \exp\{- \int_{-\infty}^{t}h(u)du\}$ and hence $P(\epsilon > t) = \exp\{- \int_{t -\Delta}^{t}h(u)du\}P(\epsilon > t - \Delta)$, where $h(t)$ is the hazard function of $\epsilon$. Thus,
Assumption \ref{ass: tail prob} is satisfied if $h(t) \to \infty$ as $t\to \infty$. Specifically, Assumption \ref{ass: tail prob} can be satisfied by the Gompertz distribution, which is suitable for modeling extreme events in hydrology, and the generalized gamma distribution with a shape parameter larger than the scale parameter, which is widely adopted in survival analysis \citep{cox2007parametric}.

Assumption \ref{ass: tail prob} is about the upper tail of the error term $\epsilon$. It's possible that the light-tailedness is satisfied by the lower tail.
As a counterpart to  Assumption \ref{ass: tail prob}, an analogous version on the lower tail of $\epsilon$ is also sufficient to identify the causal effect. 
In this section, we focus on identification under Assumption \ref{ass: tail prob}. See Supplementary Material Section \ref{app: select tail} for more discussions on the general case where either the lower or upper tail of $\epsilon$ satisfies the light-tailedness condition and researchers don't know which tail is light.

Assumption \ref{ass: balance} is a regularity condition about the discrepancy between the conditional distribution of $\epsilon$ given $X$ and its marginal distribution. It imposes restrictions on the dependence between $\epsilon$ on $X$.
It holds naturally when $X$ is exogenous, i.e., $X \Perp \epsilon$. 
When $X$ is endogenous, it can be satisfied in many familiar situations as illustrated in the following examples.
In the first example, the correlation between $X$ and $\epsilon$ is driven by an unmeasured confounder.
\begin{example}[Unmeasured confounder]\label{eg: unmeasured confounder}
	Suppose $X$ is correlated with $\epsilon$ and the latent unconfoundedness $\epsilon \Perp X\mid U$ holds, where $U$ is a vector of unmeasured confounders. The following causal diagram provides an illustration for this example.
	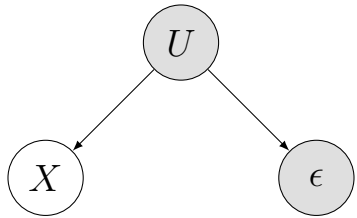
\begin{figure}[h]
		\centering
		\begin{tikzpicture}[scale = 0.6]
			\node [circle, draw=black, fill=white, inner sep=3pt, minimum size=1cm] (x) at (0,0) {\large $X$};
			\node [obs, minimum size=1cm] (e) at (6,0) {\large $\epsilon$};
			\node [obs, minimum size=1cm] (u) at (3,3.0) {\large $U$};
			\path [draw,->] (u) edge (x);
			\path [draw,->] (u) edge (e);
		\end{tikzpicture}
		\caption{An illustration for the endogeneity induced by an unmeasured confounder.}
	\end{figure}
	Then, Assumption \ref{ass: tail prob} can be satisfied if $\epsilon$ follows one of the common distributions mentioned in the paragraph after Assumption \ref{ass: balance}. Assumption \ref{ass: balance} can be satisfied if the distribution of $U$ is not extremely imbalanced in each exposure level, i.e., 
	\begin{equation}\label{eq: balance confounder}
		c_{x} \leq \frac{p_{U\mid X}(u\mid x)}{p_{U}(u)} \leq C_{x}
	\end{equation}
	for any $u$, some $0 < c_{x} < C_{x}$, and any $x\in\cX$. If $U\in\{1, \dots, K\}$ is a categorical variable, a sufficient condition for \eqref{eq: balance confounder} is $p_{U\mid X}(u\mid x) > 0$ for any $u\in\{1,\dots, K\}$ and $x\in \cX$.
	For a binary exposure $X \in \{0, 1\}$,  a sufficient condition for \eqref{eq: balance confounder} is
	\begin{equation}\label{eq: overlap}
		c \leq P(X = 1\mid U = u) \leq 1 - c,
	\end{equation}
	for some constant $0 < c < 1$. 
	Inequality \eqref{eq: overlap} is the strong overlap condition that is commonly adopted in causal inference and sensitivity analysis \citep{rosenbaum1987sensitivity, rothe2017robust, zhang2021selecting}. 
\end{example} 

\begin{example}[Selection bias]\label{eg: selection bias}
	Suppose the observed data is a biased sampling from the model $Y^*= f_{*}(X^*) + \epsilon^{*}$ with $X^{*}\Perp \epsilon^{*}$. 
	Let $S$ be the sampling indicator and $Y=SY^{*},X=SX^{*}$. Conditional on $S = 1$, the observed data  $(Y, X)$  follow \eqref{eq: additive model} with $f_{0}$ and $\epsilon$ being constant shifts from  $f_{*}$ and $\epsilon^{*}$, respectively.  
	If  $S$  depends on both $X^{*}$ and $Y^{*}$, $X^{*}$ and $\epsilon^{*}$ can be dependent conditional on $S = 1$. Hence, the selection can lead to dependence between $X$ and $\epsilon$ in the observed data. See the following diagram for an illustration.
	\begin{figure}[h]
		\centering
		\begin{tikzpicture}[scale = 0.6]
			\node [circle, draw=black, fill=white, inner sep=3pt, minimum size=1cm] (x) at (0,0) {\large $X^{*}$};
			\node [obs, minimum size=1cm] (e) at (6,0) {\large $\epsilon^{*}$};
			\node [circle, draw=black, fill=white, inner sep=3pt, minimum size=1cm] (s) at (3,3.0) {\large $S$};
			\path [draw,->] (x) edge (s);
			\path [draw,->] (e) edge (s);
		\end{tikzpicture}
		\caption{An illustration for the endogeneity induced by selection.}
	\end{figure}
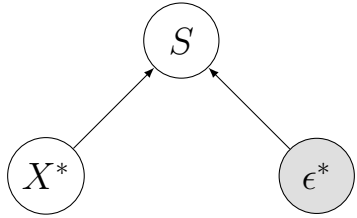
	
	In this example, Assumption \ref{ass: balance} is satisfied conditional on $S = 1$ if the selection probability $P(S = 1\mid X^{*}, Y^{*}) \geq c$ for some constant $0 < c < 1$, i.e., bounded away from zero. Under the above assumption of the selection probability, Assumption \ref{ass: tail prob} is satisfied by the error $\epsilon$ conditional on $S = 1$ if it is satisfied by the original error $\epsilon^{*}$ in the absence of selection bias. 
\end{example}  

\begin{example}[Measurement error]\label{eg: measurement error}
	Assume $Y = W^{\T}\theta_{0} + V$ where $W\Perp V$ and $\theta_{0}$ represent the causal effect of the exposure $W$ on the outcome $Y$. In practice, the true exposure may not be observable. Instead, an error-contaminated exposure $X = W + U$ is observed where $U\Perp (W, V)$ is the measurement error. Then, we have 
	$Y = X^{\T}\theta_{0} + \epsilon$ with $\epsilon = - U^{\T}\theta_{0} + V$, and $\theta_{0}$ can be estimated by considering this linear model between $Y$ and $X$. However, estimation under model $Y = X^{\T}\theta_{0} + \epsilon$ faces the endogeneity problem because the measurement error $U$ introduces dependence between $X$  and $\epsilon$. In this example, Assumption \ref{ass: tail prob} is satisfied if $- U^{\T}\theta_{0} + V$ follows one of the common distributions mentioned in the paragraph after Assumption \ref{ass: balance}, e.g., if $U$ and $V$ are jointly normal.
	Assumption \ref{ass: balance} is satisfied if the density of the true exposure satisfies $c < p_{W}(w) < C$ for some constants $0 < c < C$.
\end{example}

We next establish the identification of the causal effect under Assumptions \ref{ass: tail prob} and \ref{ass: balance}.
For any $\tau \in [0, 1]$ and any random variable $G$, let $q_{G}(\tau)$ be the marginal $(1 - \tau)$-quantile of $G$ and $q_{G}(x; \tau)$ the $(1 - \tau)$-quantile of $G$ conditional on $X = x$. 
The following proposition characterizing the behavior of the error term at the extreme quantile is the key result that motivates our identification strategy.

\begin{proposition}\label{prop: quantile homo}
	Under Assumptions \ref{ass: tail prob} and \ref{ass: balance}, for any $x \in \cX$, we have $|q_{\epsilon}(x; \tau) - q_{\epsilon}(\tau)| \to 0$ as $\tau \to 0$.
\end{proposition}

Proposition \ref{prop: quantile homo} shows that, although the conditional distribution of $\epsilon$  given $X = x$ depends on $x$, the dependence vanishes at the extreme quantile under Assumptions \ref{ass: tail prob} and \ref{ass: balance}. 
This suggests that the relationship between $X$ and  $Y$  is approximately unaffected by the endogeneity at the extreme quantiles, although the endogeneity is generally non-negligible. This motivates us to identify the causal effect by leveraging extreme quantiles.

Note that $q_{Y}(x; \tau) = f_{0}(x) + q_{\epsilon}(x; \tau)$ for any $\tau \in (0, 1)$ and $x\in\cX$. For any reference level $x_{0}$, we have $q_{Y}(x; \tau) - q_{Y}(x_{0}; \tau) = f_{0}(x) - f_{0}(x_{0}) + q_{\epsilon}(x; \tau) - q_{\epsilon}(x_{0}; \tau)$. Proposition \ref{prop: quantile homo} implies $\lim_{\tau \to 0}\{q_{\epsilon}(x; \tau) - q_{\epsilon}(x_{0}; \tau)\} \to 0$. Hence, we have $\theta(x, x_{0}) = f_{0}(x) - f_{0}(x_{0}) = \lim_{\tau \to 0}\{q_{Y}(x; \tau) - q_{Y}(x_{0}; \tau)\}$, which identifies $\theta(x, x_{0})$. 
Thus, we obtain the following identification result based on the extreme quantiles of $Y$.

\begin{theorem}\label{thm: identification}
	Under model \eqref{eq: additive model} and Assumptions \ref{ass: tail prob} and \ref{ass: balance}, for any   $x, x_{0}\in \cX$, we have $\theta(x, x_{0}) = \lim_{\tau \to 0}\{q_{Y}(x; \tau) - q_{Y}(x_{0}; \tau)\}$.
\end{theorem}

Theorem \ref{thm: identification} provides a formal justification of identifying the causal effect with extreme quantiles, by noting that the quantile $q_{Y}(x; \tau)$ can be obtained from the joint distribution of $(X, Y)$. Intuitively, this can be achieved because, at the extreme, the relationship between $X$ and $Y$ is predominantly influenced by the causal connection rather than the random error if the error term is light-tailed. 
In the absence of the exogeneity assumption or auxiliary variables, Theorem \ref{thm: identification} directly identifies the causal effect $\theta(x, x_{0})$ without explicit identification of the structural function $f_{0}$. This stands in contrast to conventional identification strategies such as the exogeneity or IV-based methods, which identify the entire structural function $f_{0}$ and then identify the causal effect by comparing its values at different exposure levels.

The rationale behind Theorem \ref{thm: identification} is best illustrated with a bounded error example.
Suppose $X$ is a scalar, $X = U + \eta_{X}$, and $\epsilon = 3U + \eta_{\epsilon}$, where $\eta_{X}$, $\eta_{\epsilon}$, and $U$ are mutually independent and $U$ takes the values
$\{-1, 1\}$ equiprobably. Assume for simplicity that $\eta_{X}\sim N(0, 9)$ and $\eta_{\epsilon} \sim U(-3, 3)$. The logic of subsequent derivations also applies to the important case where both $\eta_{X}$ and $\eta_{\epsilon}$ follow a normal distribution despite some additional technical difficulties. Conditional on $X = x$, $\epsilon$ follows the uniform mixture distribution $(1 - \pi_{x}) U(-6, 0) + \pi_{x} U(0, 6) $, where $\pi_{x} = \phi((x - 1) / 3) / \{\phi((x + 1) / 3) + \phi((x - 1) / 3)\}$ and $\phi(\cdot)$ is the density of the standard normal distribution. Both Assumptions \ref{ass: tail prob} and \ref{ass: balance} are met in this example. Next, we demonstrate how the result of Theorem \ref{thm: identification} is established in this example. Note that the component $\pi_{x} U(0, 6)$ contributes all of the probability mass in the upper tail of the uniform mixture distribution $(1 - \pi_{x}) U(-6, 0) + \pi_{x} U(0, 6)$. Thus, under model \eqref{eq: additive model}, for $\tau$ close to zero, we have $q_{Y}(x; \tau) = f_{0}(x) + 6 - 6\tau / \pi_{x}$; and similarly, $q_{Y}(x_{0}; \tau) =  f_{0}(x_{0}) + 6 - 6\tau / \pi_{x_{0}}$ for any reference level $x_{0}$. Consequently, the causal effect $\theta(x, x_{0}) = f_{0}(x) - f_{0}(x_{0}) = q_{Y}(x; \tau) - q_{Y}(x_{0}; \tau) + 6 \tau (\pi_{x}^{-1} - \pi_{x_{0}}^{-1}) \approx q_{Y}(x; \tau) - q_{Y}(x_{0}; \tau)$  when $\tau$ is small, which effectively substantiates the core assertion of Theorem \ref{thm: identification}. 

\begin{figure}[h]
	\centering
	\subfigure[]{\includegraphics[scale = 0.16]{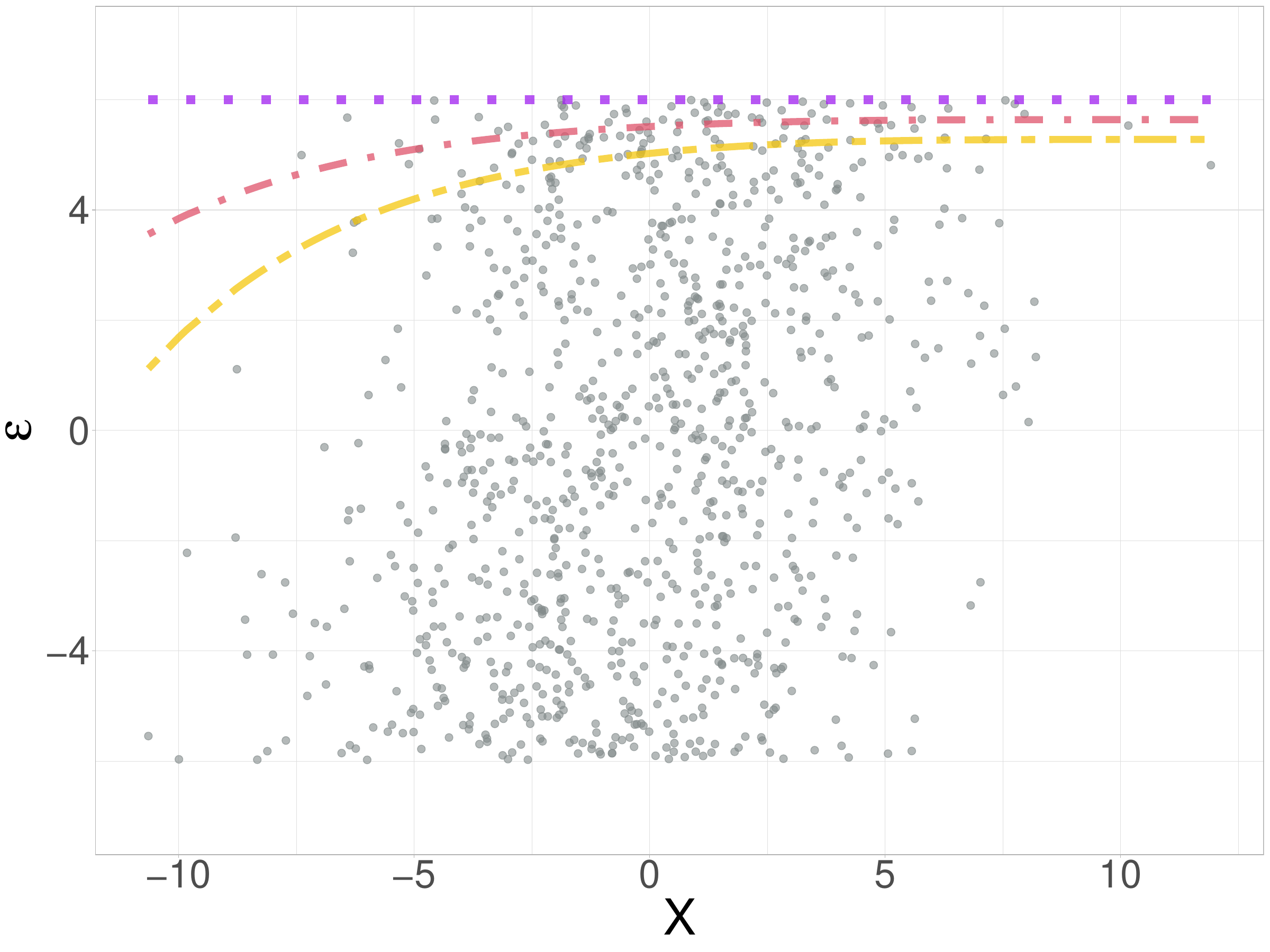}
	}
	\subfigure[]{\includegraphics[scale = 0.16]{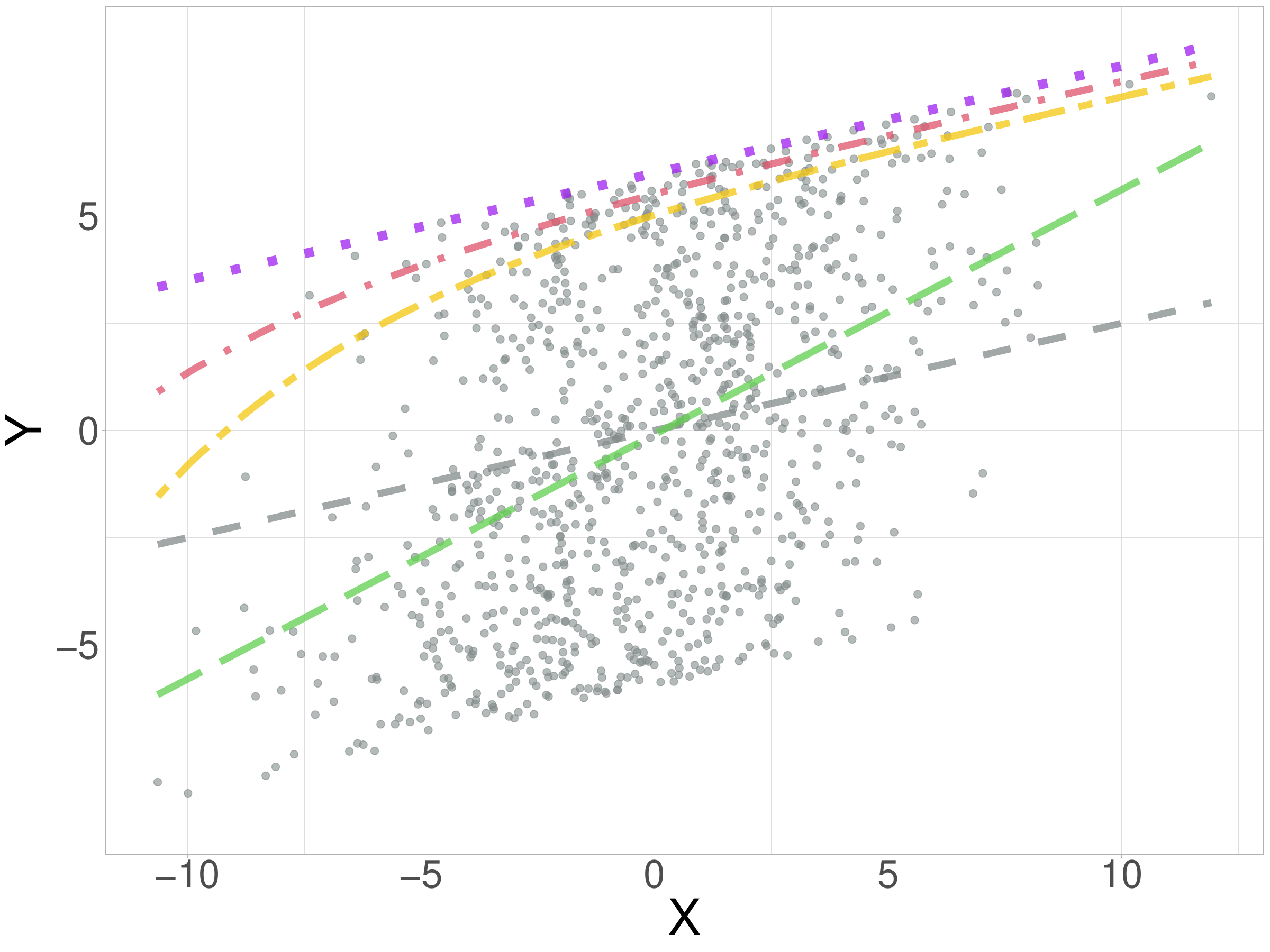}
	}
	\caption{An illustration of the identification result in the above example with $f_{0}(x) = x / 4$. Grey points in (a): realizations of $(X, \epsilon)$; Grey points in (b): realizations of $(X, Y)$; grey dashed line: the true $f_{0}(x)$; green longdash: least squares regression curve; yellow twodash in (a): $q_{\epsilon}(x; 0.06)$; yellow twodash in (b): $q_{Y}(x; 0.06)$; red dashed-dotted in (a): $q_{\epsilon}(x; 0.03)$;
		red dashed-dotted in (b): $q_{Y}(x; 0.03)$; purple dotted line in (a): $\lim_{\tau \to 0}q_{\epsilon}(x; \tau)$; purple dotted line in (b): $\lim_{\tau \to 0}q_{Y}(x; \tau)$.}\label{fig: illustration identification}
\end{figure}

Figure~\ref{fig: illustration identification} illustrates the identification result in the above example. Two quantile curves with $\tau = 0.06$ and $\tau = 0.03$ are plotted in Figure~\ref{fig: illustration identification} for illustration. The limiting curve $\lim_{\tau \to 0}q_{\epsilon}(x; \tau)$ in Fig.~\ref{fig: illustration identification} (a) represents the upper boundary of $\epsilon$'s support conditional on $X = x$, which is independent of $x$ because the support of $\epsilon$ does not depend on $X$.  This observation verifies the claim in Proposition \ref{prop: quantile homo} for this specific example. The least squares regression curve in Fig.~\ref{fig: illustration identification} (b) is not close to the true $f_{0}(x)$ due to the presence of endogeneity. On the other hand, the curve $\lim_{\tau \to 0}q_{Y}(x; \tau)$ in Fig.~\ref{fig: illustration identification} (b) is parallel to $f_{0}(x)$. This phenomenon implies that the causal effect $\theta(x, x_{0}) = f(x) - f(x_{0}) = \lim_{\tau \to 0}q_{Y}(x; \tau) - \lim_{\tau \to 0}q_{Y}(x_{0}; \tau)  = \lim_{\tau \to 0}\{q_{Y}(x; \tau) - q_{Y}(x_{0}; \tau)\}$ and demonstrates the identification result in Theorem \ref{thm: identification}. The limit $\lim_{\tau \to 0}q_{Y}(x; \tau)$ is not well-defined when $Y$ is unbounded. However, Theorem \ref{thm: identification} shows that the relationship $\theta(x, x_{0}) = \lim_{\tau \to 0}\{q_{Y}(x; \tau) - q_{Y}(x_{0}; \tau)\}$ remains true when $Y$ is unbounded as long as Assumptions \ref{ass: tail prob} and \ref{ass: balance} holds.

The utility of extreme values in identification problems has previously been demonstrated in identifying the sample selection model \citep{d2013another,d2018extremal} and the causal diagram under a linear structural equation model \citep{gnecco2021causal}. 
Our approach extends the application of extreme values to the identification of causal effects within a general endogenous additive structural equation, encompassing a broad array of important applications. 
While employing extreme values for causal effect identification is not unprecedented, our study is distinguished by addressing a broader problem and introducing novel identification assumptions and methodologies that markedly differ from existing approaches in the literature.
Specifically, \cite{d2013another} and \cite{d2018extremal} identify the causal effect under the sample selection model in Example \ref{eg: selection bias} based on the assumption that
\begin{equation}\label{ass: independence at infinity}
	\lim_{y\to y_{U}}P(S=1\mid X^{*}=x, Y^{*} = y) = l,
\end{equation}	
where $l$ is a constant independent of $x$, and $y_{U}$ is the upper bound of the support of $Y$ (can be infinity). 
In contrast, Theorem \ref{thm: identification} can identify the causal effect under the sample selection model and the mild condition that $P(S=1\mid X^{*}, Y^{*}) \geq c$ for some $0 < c < 1$ when the error term is light-tailed. In addition, the identification results in \cite{d2013another} and \cite{d2018extremal} are limited to the sample selection model, while Theorem \ref{thm: identification} also applies to other important problems including the estimation problems with unmeasured confounders or measurement errors.
In a related vein, \cite{gnecco2021causal} consider the causal discovery problem in linear models with heavy-tailed error terms. 
Their approach can detect the existence of the causal effect by utilizing extreme values under certain conditions, but cannot identify the effect size. 
Our method does not require the linearity assumption and can identify the effect size.

\begin{remark}
	One may notice that Assumption \ref{ass: balance} excludes the case where $\epsilon$ and $X$ are correlated and jointly normal. However, in the presence of endogeneity, our model \eqref{eq: additive model} allows $f_{0}$ to be nonlinear and Assumption \ref{ass: tail prob} allows $\epsilon$ and components of $X$ to be marginally normal, which is distinct from the linear non-Gaussian assumption in the causal discovery literature \citep{hoyer2008estimation, shimizu2011directlingam}.
\end{remark}

\section{Estimation and Inference}
\subsection{Nonparametric Estimation}
After establishing the identification of the causal effect, in this section, we consider the estimation of the causal effect $\theta(x, x_{0})$ based on $n$ independent and identically distributed (i.i.d.) observations $(X_{1}, Y_{1}),\dots,(X_{n}, Y_{n})$. 
According to Theorem \ref{thm: identification}, one can estimate the causal effect by estimating the conditional quantile function $q_{Y}(x; \tau)$ for some small $\tau$. 
Let $v(x)=(v_{1}(x), \dots, v_{p}(x))^\T$ be a vector of basis functions with $v_{1}(x) \equiv 1$.
We use a series approximation $q_{Y}(x; \tau) \approx v(x)^{\T}\beta_{\tau}$ to construct the estimator.
Let $\tau$ be a small positive number. Then, the extreme-based estimator is constructed as follows:
\begin{itemize}
	\item For any $x \in \cX$, estimate the conditional quantile $q_{Y}(x; \tau)$ by $v(x)^{\T}\hbeta$, where
	\[
	\hbeta = \mathop{\arg\min}_{\beta} \frac{1}{n}\sum_{i=1}^{n}\rho_{1 - \tau}(Y_{i} - v(X_{i})^{\T}\beta)
	\]
	is obtained from the quantile regression and $\rho_{1 - \tau}(z) = z(1 - \tau-1\{z < 0\})$ is the check function;
	\item Define the extreme-based estimator as
	\[
	\hat{\theta}(x, x_{0}) = \{v(x) - v(x_{0})\}^{\T}\hbeta.
	\]
\end{itemize}
The estimator $\hbeta$ is the quantile regression estimator which can be calculated utilizing standard quantile regression algorithms. It is worth mentioning that quantile regression is not the only way to estimate the causal effect based on our identification strategy. For example, let ${\rm ES}_{Y}(x; \tau) = \tau^{-1}\int_{0}^{\tau}q_{Y}(x; t)dt$ be the $\tau$-upper expected shortfall of $Y$ conditional on $X = x$. Then, according to the same arguments in the proofs of Proposition \ref{prop: quantile homo} and Theorem \ref{thm: identification}, it can be shown that $\theta(x, x_{0}) = \lim_{\tau \to 0}\{{\rm ES}_{Y}(x; \tau) - {\rm ES}_{Y}(x_{0}; \tau)\}$ under Assumptions \ref{ass: tail prob} and \ref{ass: balance}. Thus, one can also estimate the causal effect using the expected shortfall regression \citep{girard2021extreme}. Another potential approach is expectile regression, which is closely related to expected shortfall regression \citep{taylor2008estimating}. To be specific, we focus on estimation through the quantile regression in this paper. Investigations on the optimal strategy will be studied elsewhere.

For any function $g$, let $\|g\|_{\infty} = \sup_{x\in \cX}|g(x)|$ be the infinity norm. Let $\barbeta$ be a minimum point of $\|q_{Y}(\cdot; \tau) - v(\cdot)^{\T}\beta\|_{\infty}$ over all $\beta$. Then $v(\cdot)^{\T}\barbeta$ is the optimal approximation of $q_{Y}(\cdot; \tau)$ in the space spanned by $\{v_{j}\}_{j=1}^{p}$. Let $\zeta_{a} = \|q_{Y}(\cdot; \tau) - v(\cdot)^{\T}\barbeta\|_{\infty}$ be the approximation error, $\kappa_{\infty} = \sup_{\beta}\|v(\cdot)^{\T}\beta\|_{\infty} / \|\beta\|_{\Sigma}$, and $\Sigma = E\{v(X)v(X)^{\T}\}$. For any $p$-dimensional vector $\delta$, let $\|\delta\|_{\Sigma} = \sqrt{\delta^{\T}\Sigma\delta}$. To establish the consistency of the proposed extreme-based estimator, we invoke the following conditions. 
\begin{condition}[Liptchitz continuity]\label{cond: Lip}
	There is some constant $C_{L}$ such that $|f_{Y\mid X}(y_{1}\mid x) - f_{Y\mid X}(y_{2}\mid x)| \leq C_{L}|y_{1} - y_{2}|$ for any $y_{1}, y_{2}\in \cY$ and $x \in \cX$.
\end{condition}

\begin{condition}\label{cond: bound dens}
	There is some constant $C_{f}$ such that $f_{Y\mid X}(y\mid x) \leq C_{f}$ for any $y\in \cY$ and $x \in \cX$.
\end{condition}

\begin{condition}\label{cond: bound harzard}
	For any $\tau \in (0, 1)$, there is some constant $c_{\tau}$ such that $f_{Y\mid X}(q_{Y}(\tau; x)\mid x) \geq c_{\tau}$ for any $x \in \cX$.
\end{condition}
\begin{condition}\label{cond: rate consistency}
	$\tau \to 0$, $\zeta_{a}/\tau^{2} \to 0$, $\max\{\kappa_{\infty}, \tau^{-1}\}\sqrt{p\log n / (n\tau)}\to 0$, and $c_{\tau} \geq c\tau$ for some constant $c > 0$.
\end{condition}
Conditions \ref{cond: Lip} and \ref{cond: bound dens} are mild regularity conditions on the conditional density $f_{Y\mid X}$, which require the conditional density to be Liptchitz continuous and bounded, respectively. Condition \ref{cond: bound harzard} requires the conditional density to be bounded from below at the conditional quantile. All these three conditions are standard in the literature of quantile regression \citep{kato2011group, he2023smoothed}. Condition \ref{cond: rate consistency} is a conditional on the convergence rates of $\tau$, $\zeta_{a}$, $\kappa_{\infty}$, $c_{\tau}$, $p$, and $n$. The quantities in Condition \ref{cond: rate consistency} can be characterized in specific examples.
To fix the idea, assume $X$ is a $d$-dimensional continuous exposure with bounded support.  Suppose the quantile function $q_{Y}(x; \tau)$ has bounded partial derivatives up to order $s$. Then $\zeta_{a} = O(p^{-s/d})$ if the basis functions are tensor products of B-splines, trigonometric polynomial functions or wavelet bases \citep{lorentz1986approximation, chen2007large}. In addition, we have $\zeta_{a} = 0$ when the model is correctly specified in the sense that $q_{Y}(x; \tau) = v(x)^{\T}\beta_{\tau}$ for some $\beta_{\tau}$. The quantity $\kappa_{\infty}$ is a measure of irregularity of the finite-dimensional linear space spanned by the basis functions $\{v_{j}\}_{j=1}^{p}$. We have $\kappa_{\infty} = O(\sqrt{p})$ for the above basis functions. Moreover, we have $c_{\tau} \geq c\tau$ for any $\tau$ and some constant $c > 0$ provided the hazard function of $\epsilon$ is bounded away from zero. 

To obtain the uniform consistency result, the following stronger version of Assumption \ref{ass: balance} is required.
\begin{assumption}\label{ass: balance strong}
	There are some constants $0 < c < C$ such that $c \leq p_{\epsilon\mid X}(e\mid x) / p_{\epsilon}(e) \leq C$ for any $e \in \cE$ and $x\in \cX$.
\end{assumption}
Assumption \ref{ass: balance strong} additionally requires the lower and upper bounds of the density ratio to be uniform in $x$.	Although Assumption \ref{ass: balance strong} is stronger than Assumption \ref{ass: balance}, the conditions in Examples \ref{eg: unmeasured confounder}, \ref{eg: selection bias}, and \ref{eg: measurement error} are also sufficient for Assumption \ref{ass: balance strong}.

The following theorem establishes the consistency for the proposed extreme-based estimator.
\begin{theorem}\label{thm: consistency}
	Under Assumptions \ref{ass: tail prob}, \ref{ass: balance} and Conditions \ref{cond: Lip}--\ref{cond: rate consistency}, we have $|\hat{\theta}(x,x_{0}) - \theta(x, x_{0})| \to 0$ in probability for any $x\in\cX$. In addition, if Assumption \ref{ass: balance strong} also holds, then $\sup_{x\in \cX}|\hat{\theta}(x,x_{0}) - \theta(x, x_{0})| \to 0$ in probability.
\end{theorem}

Theorem \ref{thm: consistency} shows that the causal effect can be consistently estimated through extreme quantile regression in the presence of endogeneity when the error term is light-tailed. 
Theorem \ref{thm: consistency} reveals a feature of extreme quantile regression:  the invulnerability to endogeneity when the error term is light-tailed,
which has not been previously appreciated in causal inference or quantile regression to our knowledge. 
Quantile regression with endogeneity has been studied under general nonseparable models, where IVs are required to achieve identifiability \citep{chernozhukov2013quantile}. Our results contribute to this field by demonstrating that in an additive model with a light-tailed error term, the causal effect can be identified and consistently estimated through quantile regression without the assistance of IVs.

Theorem \ref{thm: consistency} can be obtained from the following non-asymptotic error bound (Lemma \ref{lemma: error bound}). Define $\kappa_{3} = \sup_{\beta}E\{(v(X)^{\T}\beta)^{3}\}/\|\beta\|_{\Sigma}^{3}$, $\zeta_{\tau}(x, x_{0}) = |q_{\epsilon}(x; \tau) - q_{\epsilon}(x_{0}; \tau)|$, 
\[
\nu_{n,p,\tau} = \max\left\{1, \log\left[1 + \sqrt{\frac{n}{2p(\tau + C_{f}\zeta_{a})}}\right]\right\},  
\nu_{a, \tau} = \max\left\{\sqrt{2}\kappa_{\infty}, \frac{6\kappa_{3} C_{f}}{(c_{\tau} - C_{L}\zeta_{a})}, \frac{18\kappa_{3} C_{L}(\tau + C_{f}\zeta_{a})}{(c_{\tau} - C_{L}\zeta_{a})^{2}}\right\}.
\]
Then, we have the following non-asymptotic result.	
\begin{lemma}\label{lemma: error bound}
	For any $u > 0$ and $0 < \tau < 1$, suppose the approximation error $\zeta_{a}$ is sufficiently small and the sample size $n$ is sufficiently large  such that $\sqrt{6\kappa_{3} C_{L}C_{f}\zeta_{a}} + C_{L}\zeta_{a} < c_{\tau}$, $2\kappa_{3}C_{f}^{2}\zeta_{a} \leq (c_{\tau} - C_{L}\zeta_{a})(\tau + C_{f}\zeta_{a})$ and $\sqrt{n} \geq \nu_{a, \tau}\sqrt{(u + p\nu_{n,p,\tau}) / (\tau + C_{f}\zeta_{a})}$. Under Conditions \ref{cond: Lip}, \ref{cond: bound dens}, and \ref{cond: bound harzard},  
	with probability at least $1 - \exp(- u)$, we have
	\[
	\|\hbeta - \barbeta\|_{\Sigma} \leq \frac{18}{c_{\tau} - C_{L}\zeta_{a}}\max\left\{\sqrt{\frac{(\tau + C_{f}\zeta_{a})(u + p\nu_{n,p,\tau})}{n}}, \frac{C_{f}\zeta_{a}}{3}\right\},
	\]
	and 
	\[
	|\hat{\theta}(x,x_{0}) - \theta(x, x_{0})| \leq 
	\frac{36\kappa_{\infty}}{c_{\tau} - C_{L}\zeta_{a}}\max\left\{\sqrt{\frac{(\tau + C_{f}\zeta_{a})(u + p\nu_{n,p,\tau})}{n}}, \frac{C_{f}\zeta_{a}}{3}\right\} + 2\zeta_{a} + 2\zeta_{\tau}(x, x_{0}),
	\]
	for any $x,x_{0}\in \cX$.
\end{lemma}	

Lemma \ref{lemma: error bound} is general and purely about the estimation error, and it is valid even if the identification conditions in Theorem \ref{thm: identification} do not hold. However, the upper bound in Lemma \ref{lemma: error bound} involves the term $\zeta_{\tau}(x, x_{0})$, whose convergence requires further assumptions. Specifically, if the conditions of Theorem \ref{thm: identification} hold, the term $\zeta_{\tau}(x, x_{0})$ converges to zero as $\tau \to 0$. The proposed extreme-based estimator is built upon the nonparametric estimation of extreme conditional quantiles. The problem of estimating extreme conditional quantiles nonparametrically has been studied in the literature over the past decades. Several estimators have been proposed and justified asymptotically \citep{kurisu2023subsampling}. Lemma \ref{lemma: error bound} contributes to this body of work by providing a non-asymptotic error bound, which explicitly characterizes the error terms without requiring the sample size to go to infinity. The explicit nature of this non-asymptotic result facilitates further analysis built upon it. 
Notice that $\nu_{n,p,\tau} = O(\log n)$ if $n\tau \to \infty$. Under Condition \ref{cond: rate consistency}, we have $n\tau \to \infty$ and $c_{\tau} \geq c\tau$ for any $\tau$ and some constant $c > 0$. Then, Lemma \ref{lemma: error bound} implies that $|\hat{\theta}(x,x_{0}) - \theta(x, x_{0})| \to 0$ in probability if $\tau \to 0$,  $\zeta_{a}/\tau^{2} \to 0$, and $\max\{\kappa_{\infty}, \tau^{-1}\}\sqrt{p\log n / (n\tau)}\to 0$. 
Note that $\zeta_{\tau}(x, x_{0})$ is the only term in the upper bound for $|\hat{\theta}(x,x_{0}) - \theta(x, x_{0})|$ in Lemma \ref{lemma: error bound} that depends $x$. Under Assumption \ref{ass: balance strong}, we have $\sup_{x\in\cX}|\zeta_{\tau}(x, x_{0})| \to 0$ as $\tau \to 0$ according to similar arguments as those in the proofs of Proposition \ref{prop: quantile homo} and Theorem \ref{thm: identification}. Then, we have $\sup_{x\in\cX}|\hat{\theta}(x,x_{0}) - \theta(x, x_{0})| \to 0$ in probability according to Lemma \ref{lemma: error bound}. This establishes Theorem \ref{thm: consistency}.

Theorem \ref{thm: consistency} obtains the consistency result, which can not be directly employed for statistical inference. Nevertheless, the proposed extreme-based estimator $\hat{\theta}(x, x_{0})$, as a consistent estimator, can offer valuable guidance in statistical inference. For example, it can be used as a benchmark estimator for the selection of auxiliary variables (e.g., IVs or negative controls) in biological or socioeconomic studies.
Existing auxiliary variable selection methods usually focus on IV selection and often rely on assumptions such as \emph{majority valid} or \emph{plurality valid} to identify causal effects in the presence of invalid IVs \citep{kang2016instrumental,guo2018confidence, lin2024instrumental}. 
Our extreme quantile-based method offers a novel approach that is applicable to select general auxiliary variables and relies on minimal knowledge about the candidate auxiliary variables.
We discuss the details in Supplementary Material Section \ref{app: inference invalid auxiliary}.

The previous discussions have primarily addressed scenarios where the light-tailedness condition applies to the upper tail. However, it's possible that the light-tailedness could instead be pertinent to the lower tail. The causal effect $\theta(x, x_{0})$ remains identifiable in this case. Practically, it's advisable to initially assess the plausibility of light-tailedness for both tails and select the most appropriate one for conducting inference. Please refer to Supplementary Material Section \ref{app: select tail}  of for a detailed procedure to select the tail, which is applied in all the simulation studies in this paper. To implement the proposed extreme-based method, one needs to specify a suitable tail index $\tau$. Supplementary Material Section \ref{app: select tau} introduces a data-adaptive procedure for selecting $\tau$ which might be of practical interest.

\subsection{Inference under Linear Models}\label{subsec: infer linear}
Theorem \ref{thm: consistency} establishes the uniform consistency of the proposed extreme-based estimator, which do not suffice for statistical inference. Statistical inference for extreme quantiles in general nonparametric models can be challenging, even with exogeneity, due to data sparsity at the tail of outcome distributions \citep{wang2016estimation}. This challenge is amplified by the additional complexities introduced by endogeneity in our problem.
On the other hand, in certain situations, it is possible to obtain valid asymptotic approximations that can facilitate statistical inference.
In this section, we investigate the inference of the causal effect under a linear model, 
\begin{equation}\label{eq: linear model}
	Y = \mu_{0} + X^{\T}\theta_{0} + \epsilon.
\end{equation}
Model \eqref{eq: linear model} is a special case of model \eqref{eq: additive model} with $f_{0}(X) = \mu_{0} + X^{\T}\theta_{0}$. According to Theorem \ref{thm: identification}, the causal effect $\theta_{0}$ is identifiable under Assumptions \ref{ass: tail prob} and \ref{ass: balance}. We study the inference of $\theta_{0}$ in this section. Under Assumption \ref{ass: tail prob}, $\theta_{0}$ can be estimated by a quantile regression at upper extreme quantiles. Let $\tau_{n}$ be a decreasing positive sequence that converge to zero as $n \to \infty$ and
\[
(\hat{\mu}_{n}, \hat{\theta}_{n}) = \mathop{\arg\min}_{\mu, \theta} \frac{1}{n}\sum_{i=1}^{n}\rho_{1 - \tau_{n}}(Y_{i} - \mu - X_{i}^{\T}\theta).
\]
The estimator $\hat{\theta}_{n}$ is the standard quantile regression estimator and can be calculated using routine quantile regression packages.
To make statistical inference  based on $\hat{\theta}_{n}$, 
the following conditions are required to control the bias caused by endogeneity. 
\begin{condition}\label{cond: endogeneity bias} 
	(i) $\sqrt{n\tau_{n}} \|E[\{1 - \tau_{n}^{-1}P(\epsilon > q_{\epsilon}(\tau_{n})\mid X)\}X]\| \to 0$; (ii) there is some constant $\varpi > 1$ such that $E\{|q_{\epsilon}(X; \tau_{n}) - q_{\epsilon}(\tau_{n})|\} = o\{q_{\epsilon}(\tau_{n}) - q_{\epsilon}(\varpi \tau_{n})\}$ as $n \to 0$.
\end{condition}
Proposition \ref{prop: quantile homo} implies that $|q_{\epsilon}(X; \tau_{n}) - q_{\epsilon}(\tau_{n})|$ converges to zero in probability under regularity conditions when $\epsilon$ is light-tailed.
Condition \ref{cond: endogeneity bias} further requires  $|q_{\epsilon}(X; \tau_{n}) - q_{\epsilon}(\tau_{n})|$ to converge sufficiently fast so that the confounding bias can be controlled.
It is a technical condition that can be satisfied if $q_{\epsilon}(X; \tau) = q_{\epsilon}(\tau)$ with probability one for any sufficiently small $\tau$.
Under Condition \ref{cond: endogeneity bias} and certain regularity conditions, we have the following theorem.

\begin{theorem}\label{thm: inference 
		linear model}
	Under the linear model \eqref{eq: linear model}, Condition \ref{cond: endogeneity bias}, Conditions \ref{cond: Lip} and \ref{cond: R1}--\ref{cond: limit var} in Supplementary Material Section \ref{app: regularity cond}, if $n\tau_{n} \to \infty$, then
	\[
	\frac{\sqrt{n\tau_{n}}}{q_{\epsilon}(\tau_{n}) - q_{\epsilon}(\varpi \tau_{n})}(\hat{\theta}_{n} - \theta_{0}) \to N(0, \Sigma_{0})
	\]
	in distribution as $n\to \infty$, where the form of $\Sigma_{0}$ is provided in Supplementary Material Section \ref{app: inference linear model}. 
\end{theorem}
Theorem \ref{thm: inference linear model} concerns the property of $\hat{\theta}_{n}$ under the regime where $\tau_{n} \to 0$ and $n \tau_{n} \to \infty$ and establishes the asymptotic normality. Existing results in the extreme quantile regression literature establish the asymptotic normality of the quantile regression estimator under a linear conditional quantile model \citep{chernozhukov2005extremal}. The distinction here is that Theorem \ref{thm: inference linear model} is derived under a misspecified setting. Specifically, Theorem \ref{thm: inference linear model} is established under a linear structural model with endogenous exposures, which does not imply that the conditional quantile of $Y$ is linear in $X$.
The conditions in Supplementary Material Section \ref{app: inference linear model} are regularity conditions adapted from \cite{chernozhukov2005extremal}, which are imposed on the conditional distribution of $Y$ given $X$ and the covariance among different components of $X$. The convergence rate $\{q_{\epsilon}(\tau_{n}) - q_{\epsilon}(\varpi \tau_{n})\}/\sqrt{n\tau_{n}}$ is in general not equal to the parametric rate $1 / \sqrt{n}$. 
For an illustration, suppose $P(\epsilon > t) / \exp(- t^{2}) \to c$ as $t \to \infty$ for some $c > 0$.  Then, we have $q(\tau_{n}) - q(\varpi\tau_{n}) \asymp \{\log (1/\tau_{n})\}^{-1/2}$ as $\tau_{n} \to 0$. In this case, the convergence rate of $\hat{\theta}_{n}$ is $1 / \sqrt{n\tau_{n}\log(\tau_{n}^{-1})}$, which is slower than $1 / \sqrt{n}$ as $n \to \infty$ and $\tau_{n} \to 0$. 

Although Theorem \ref{thm: inference linear model} establishes the asymptotic normality of $\hat{\theta}_{n}$, it does not directly facilitate inference for $\theta_{0}$ due to the dependency of the convergence rate $\{q_{\epsilon}(\tau_{n}) - q_{\epsilon}(\varpi \tau_{n})\}/\sqrt{n\tau_{n}}$ on the unknown distribution of $\epsilon$. Fortunately, we can establish the following consistency result for bootstrap, which can be employed for inference. Specifically, suppose $\{(X_{i}^{*}, Y_{i}^{*})\}_{i = 1}^{n}$ are drawn with replacement from $\{(X_{i}, Y_{i})\}_{i = 1}^{n}$. Define
\[
(\hat{\mu}_{n}^{*}, 
\hat{\theta}_{n}^{*}) = \mathop{\arg\min}_{\mu, \theta} \frac{1}{n}\sum_{i=1}^{n}\rho_{1 - \tau_{n}}(Y_{i}^{*} - \mu - X_{i}^{*\T}\theta). 
\]
The next theorem establishes the consistency of the bootstrap procedure.
\begin{theorem}\label{thm: bootstrap}
	Under the conditions of Theorem \ref{thm: inference linear model}, we have
	\[
	\frac{\sqrt{n\tau_{n}}}{q_{\epsilon}(\tau_{n}) - q_{\epsilon}(\varpi \tau_{n})}(\hat{\theta}_{n}^{*} - \hat{\theta}_{n}) \to N(0, \Sigma_{0})
	\]
	in distribution conditional on data with probability approaching one as $n \to \infty$, where $\Sigma_{0}$ is introduced in Theorem \ref{thm: inference linear model} whose form is provided in Supplementary Material Section \ref{app: inference linear model}.
\end{theorem}
According to Theorem \ref{thm: bootstrap}, we can construct the confidence interval for $\theta_{0}$ utilizing bootstrap.  Let $B$ be a user-specified large integer. For $b = 1,\dots, B$, draw a sample $\left\{(X_{i}^{(b)}, Y_{i}^{(b)})\right\}_{i = 1}^{n}$ with replacement from $\left\{(X_{i}, Y_{i})\right\}_{i = 1}^{n}$. Define
\[
(\hat{\mu}_{n}^{(b)}, \hat{\theta}_{n}^{(b)}) = \mathop{\arg\min}_{\mu, \theta} \frac{1}{n}\sum_{i=1}^{n}\rho_{1 - \tau_{n}}(Y_{i}^{(b)} - \mu - X_{i}^{(b)\T}\theta).
\]
Let $\hat{c}_{\alpha}$ be the $1 - \alpha$ quantile of $\big\{\|\hat{\theta}_{n}^{(b)} - \hat{\theta}_{n}\|\big\}_{b = 1}^{B}$. An asymptotically valid $1 - \alpha$ confidence set for $\theta_{0}$ is $\big\{\theta: \|\theta - \hat{\theta}_{n}\| \leq \hat{c}_{\alpha}\big\}$. Next, we construct the confidence interval for each component of $\theta_{0}$.
For $j = 1, \dots, d$, let $\theta_{0, j}$, $\hat{\theta}_{n, j}$, and $\hat{\theta}_{n, j}^{(b)}$ be the $j$-th component of $\theta_{0}$, $\hat{\theta}_{n}$, and $\hat{\theta}_{n}^{(b)}$, respectively. Denote by $\hat{c}_{\alpha, j}$ the $1 - \alpha$ quantile of $\big\{|\hat{\theta}_{n, j}^{(b)} - \hat{\theta}_{n, j}|\big\}_{b = 1}^{B}$. Then the confidence interval for $\theta_{0, j}$ is $[\hat{\theta}_{n, j} - \hat{c}_{\alpha, j}, \hat{\theta}_{n, j} + \hat{c}_{\alpha, j}]$ for $j = 1,\dots, d$.

\begin{remark}
	Theorems \ref{thm: inference linear model} and \ref{thm: bootstrap} assume that $n\tau_{n} \to \infty$, i.e., $\tau_{n}$ is an intermediate order sequence in the terminology of extreme quantile regression \citep{chernozhukov2005extremal}. The above nonparametric bootstrap procedure works when the tail index sequence is of intermediate order  \citep{d2018extremal}. However, nonparametric bootstrap might be inconsistent when the quantile index is of extreme order in the sense that $n\tau_{n}$ converges to some positive constant \citep{bickel1981some,chernozhukov2016extremal}. The extremal bootstrap specifically designed for extreme quantile regression should be employed when $\tau_{n}$ is of extreme order. Please refer to \cite{chernozhukov2016extremal} for more details about the extremal bootstrap. In our numerical experiments, we set $\tau_{n} \asymp n^{-1/4}$ and the usual nonparametric bootstrap works well.
\end{remark}

\section{Simulation Study}\label{sec: sim}
In this section, we evaluate the performance of the proposed method through simulation studies.
The exposure $X$ and outcome $Y$ are generated from the following model,
\begin{equation*}
	\begin{gathered}
		X = 0.2U^{\T}\gamma_{U} + \eta_{X},\quad  \eta_{X} \sim N(0, 1),
		\quad \gamma_{U}=(d_{U}^{1/2},\ldots, d_{U}^{1/2}),\\
		Y = X\theta_{0} + 4U^{\T}\gamma_{U} + \eta_{\epsilon},\quad \eta_{\epsilon} \sim N(0, 0.25),\quad \theta_{0} = 0.4,
	\end{gathered}
\end{equation*}
where  $U$ is a $d_{U}$-dimensional unmeasured confounder with components independently following  ${\rm Binomial}(2, 0.3)$,
and the causal effect is captured by $\theta_{0} = 0.4$.

\begin{figure}[h]
	\centering
	\begin{tikzpicture}[scale = 0.6]
		\node [circle, draw=black, fill=white, inner sep=3pt, minimum size=0.5cm] (x) at (0,0) {\large $X$};
		\node [circle,draw=black,fill=white,inner sep=3pt,minimum size=0.5cm] (y) at (6,0) {\large $Y$};
		\node [obs, minimum size=0.7cm] (u) at (3,3.0) {\large $U$};
		\path [draw,->] (x) edge node[ anchor=center, above, pos=0.5,font=\bfseries] {$\theta_{0}$} (y);
		\path [draw,->] (u) edge (x);
		\path [draw,->] (u) edge (y);
	\end{tikzpicture}
	\caption{An illustration of the data generating process.}
\end{figure}
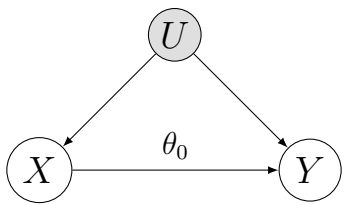

The data generating mechanism leads to the reduced form structural equation $Y = X\theta_{0} + \epsilon$ with $\epsilon = 4U^{\T}\gamma_{U} + \eta_{\epsilon}$, which is correlated with $X$. 
We implement the ordinary least squares (OLS) estimator that regresses $Y$ on $X$ and the extreme-based estimator $\hat{\theta}_{n}$ to estimate $\theta_{0}$. In addition, we construct the confidence interval for $\theta_{0}$ using the bootstrap method proposed in Section \ref{subsec: infer linear}.
The quantile index $\tau_{n}$ is set to be $0.01/n^{1/4}$ in the implementation of $\hat{\theta}_{n}$ throughout the simulation and real data analysis. 

\begin{figure}[h]
	\centering
	\subfigure[Bias, $n = 1000$]{\includegraphics[scale = 0.18]{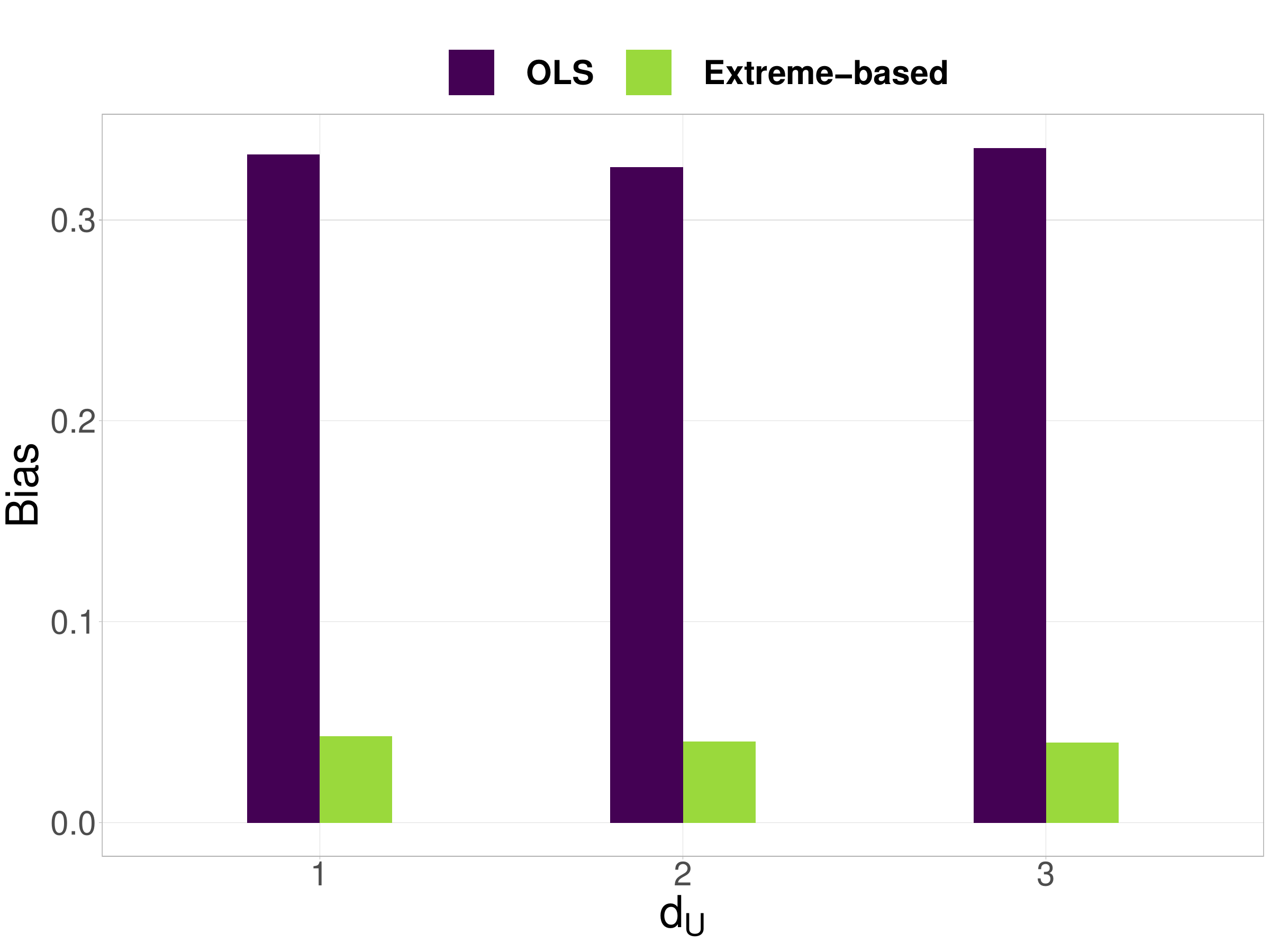}}
	\subfigure[MSE, $n = 1000$]{\includegraphics[scale = 0.18]{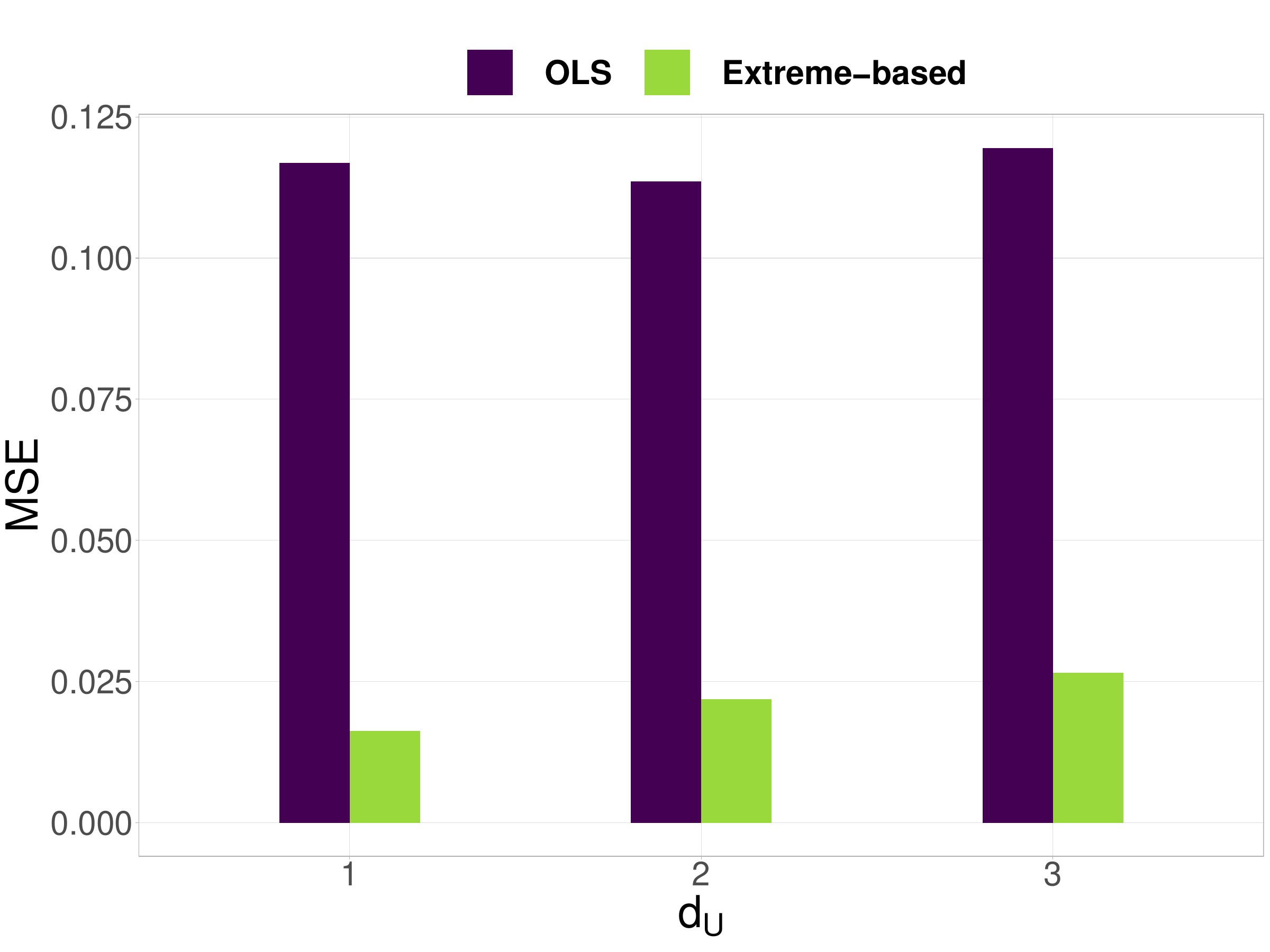}}
	\subfigure[Bias, $n = 5000$]{\includegraphics[scale = 0.18]{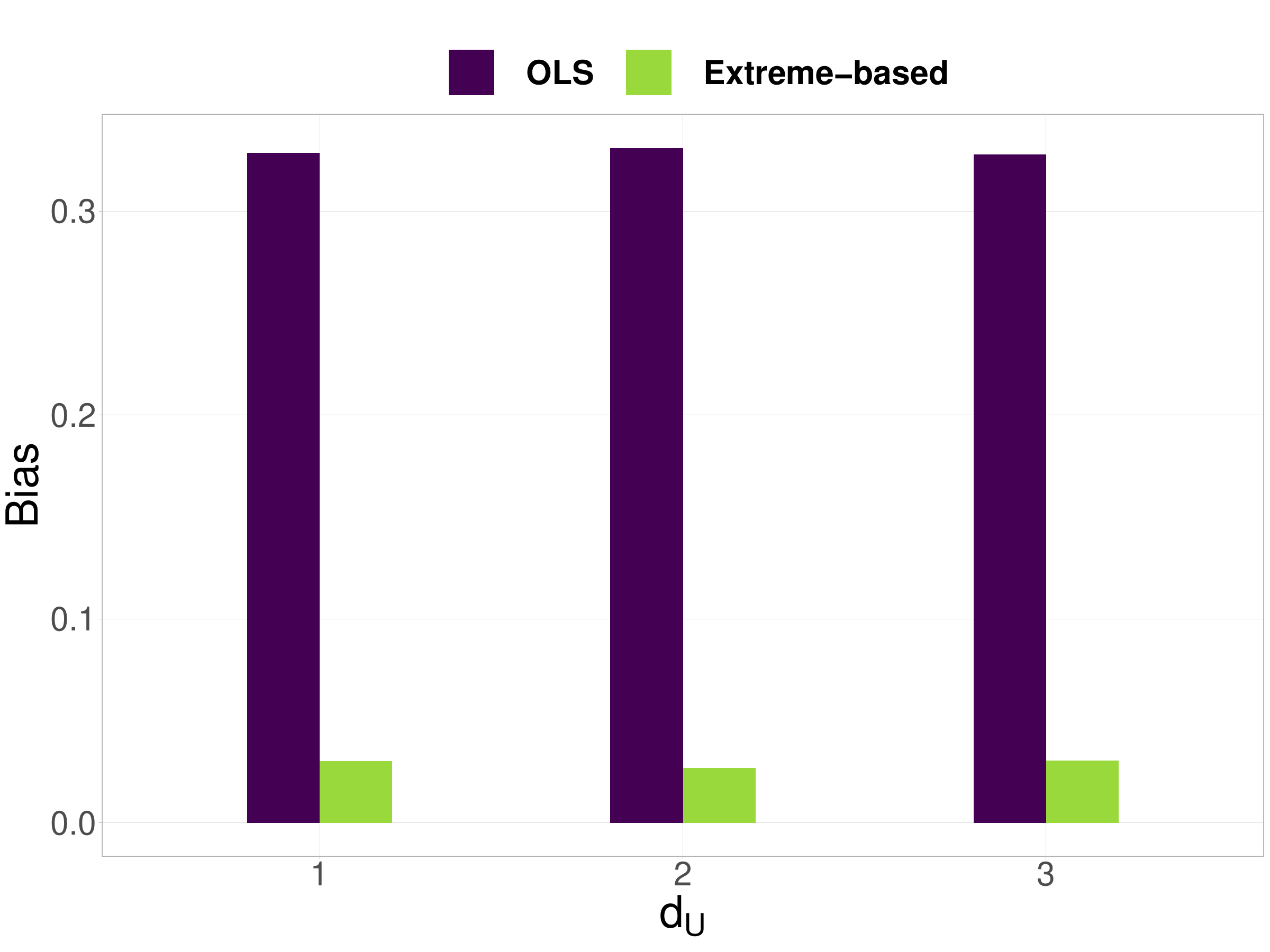}}
	\subfigure[MSE, $n = 5000$]{\includegraphics[scale = 0.18]{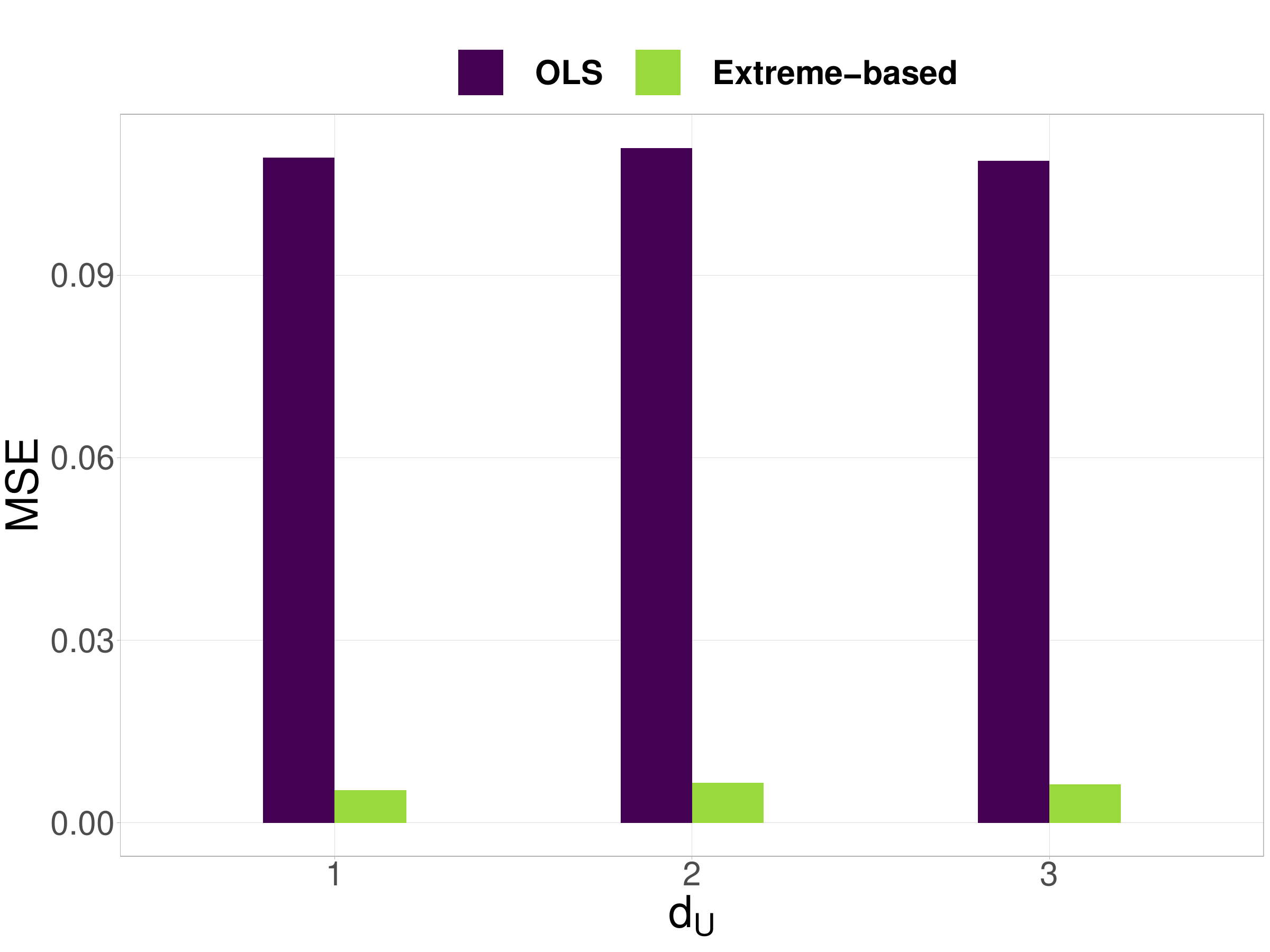}}
	\caption{Bias and MSE under the linear model with different $d_{U}$ and $n$.}\label{fig: estimation}
\end{figure}

\begin{figure}[h]
	\centering
	\subfigure[$n = 1000$]{\includegraphics[scale = 0.18]{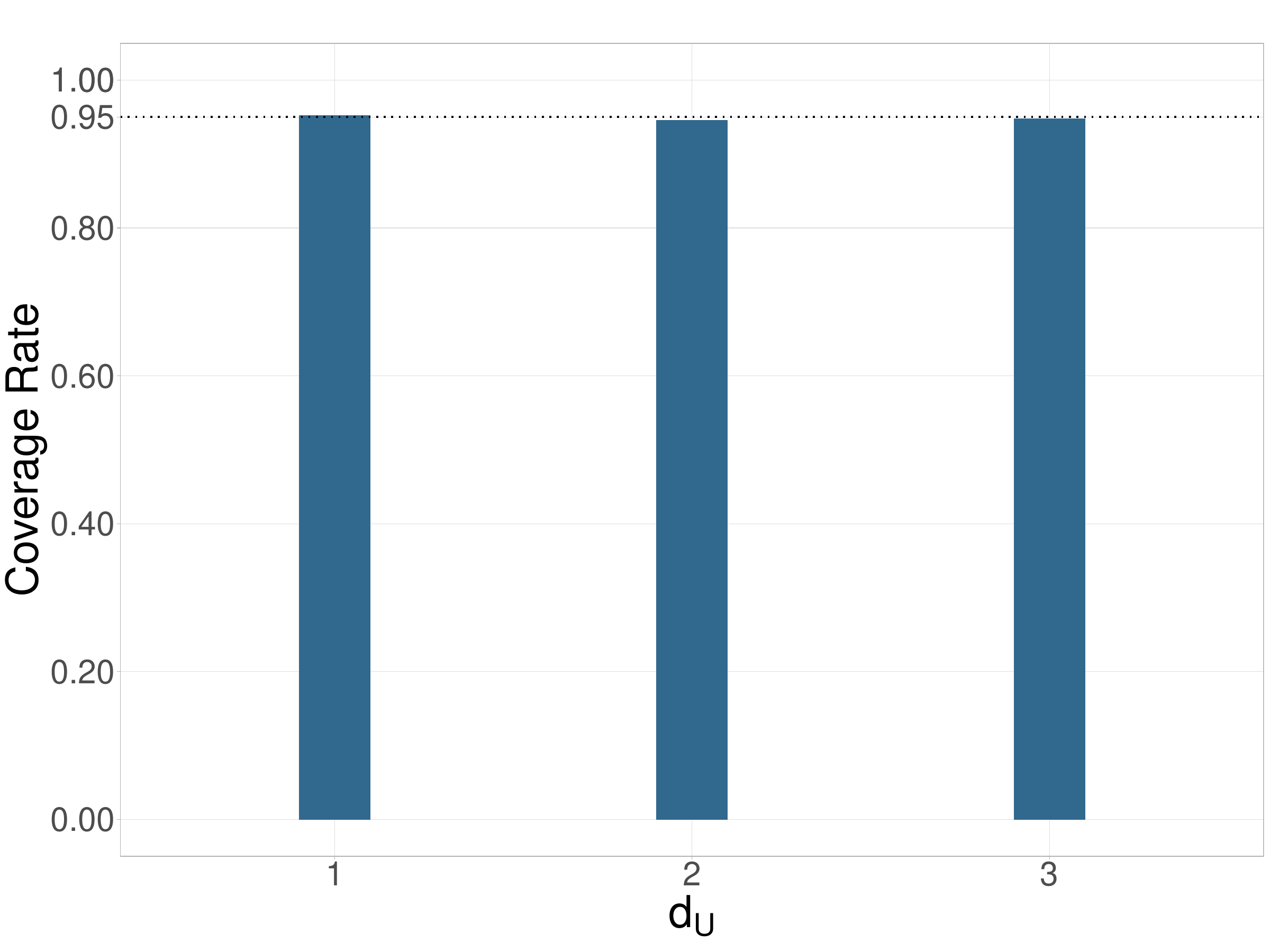}}
	\subfigure[$n = 5000$]{\includegraphics[scale = 0.18]{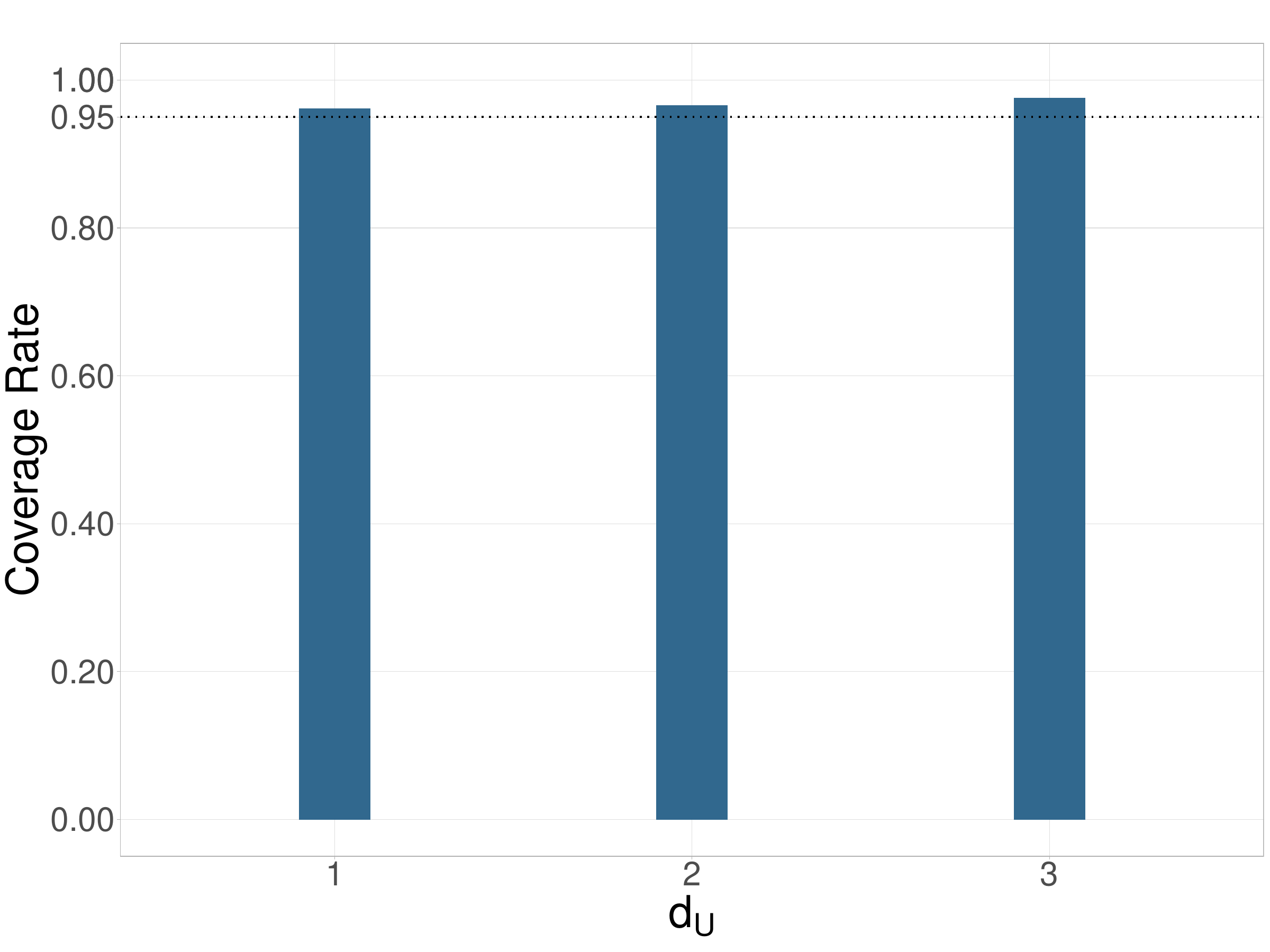}}
	\caption{Coverage rate of the $95\%$ confidence interval under the linear model with different $d_{U}$ and $n$.}\label{fig: CI}
\end{figure}

We replicate $500$ simulations at sample sizes $1000$ and $5000$, respectively.
Figure \ref{fig: estimation} shows the biases and mean square errors (MSEs) of the OLS estimator and the extreme-based estimator under different $d_{U}$ and $n$.
From Fig.~\ref{fig: estimation}, the OLS estimator has a large bias and MSE due to the endogeneity. 
The bias does not decrease as the sample size increases. 
In contrast, the extreme-based estimator has a much smaller bias and MSE in all settings, and the bias and MSE  decrease as the sample size increases. 
Figure \ref{fig: CI} shows the coverage rate of the bootstrap confidence intervals based on the extreme-based estimator. 
The coverage rates are close to the nominal level of $0.95$ across different combinations of $d_{U}$ and $n$.
These results suggest that the extreme-based method can effectively adjust for the endogeneity, with accurate point estimation and confidence interval for the causal effect.

\begin{figure}[h]
	\centering
	\subfigure[Bias, $n = 1000$]{\includegraphics[scale = 0.18]{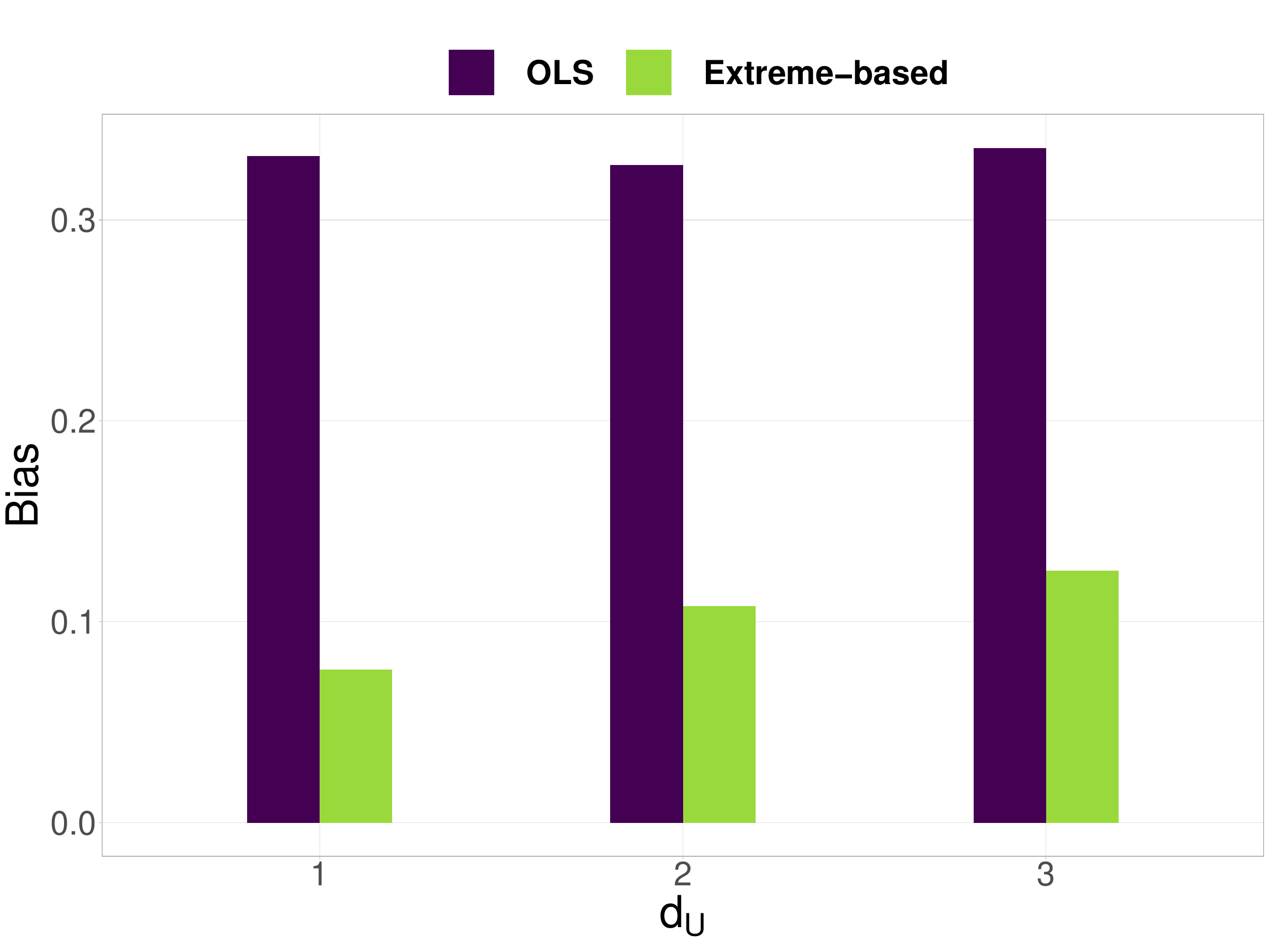}}
	\subfigure[MSE, $n = 1000$]{\includegraphics[scale = 0.18]{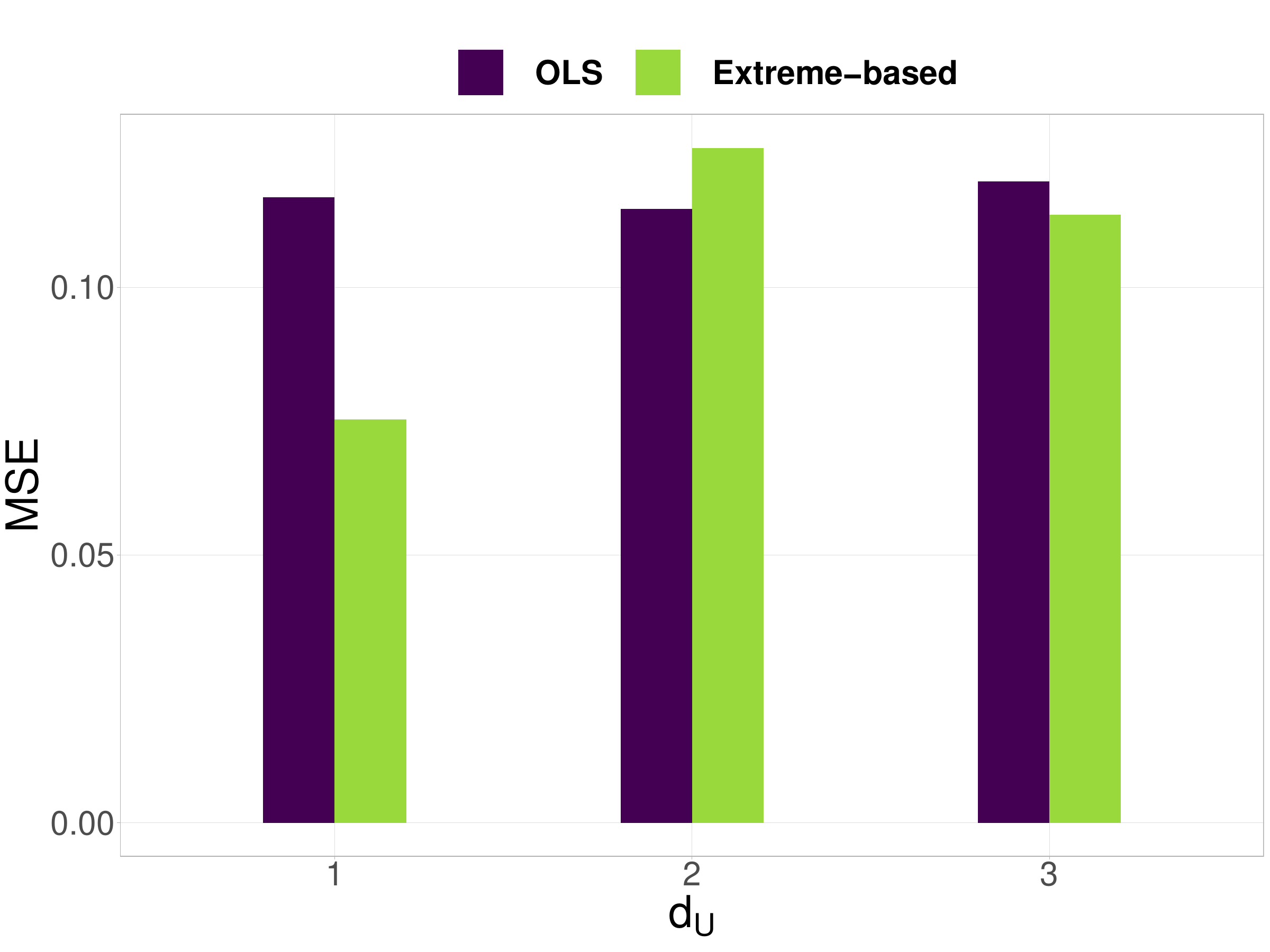}}
	\subfigure[Bias, $n = 5000$]{\includegraphics[scale = 0.18]{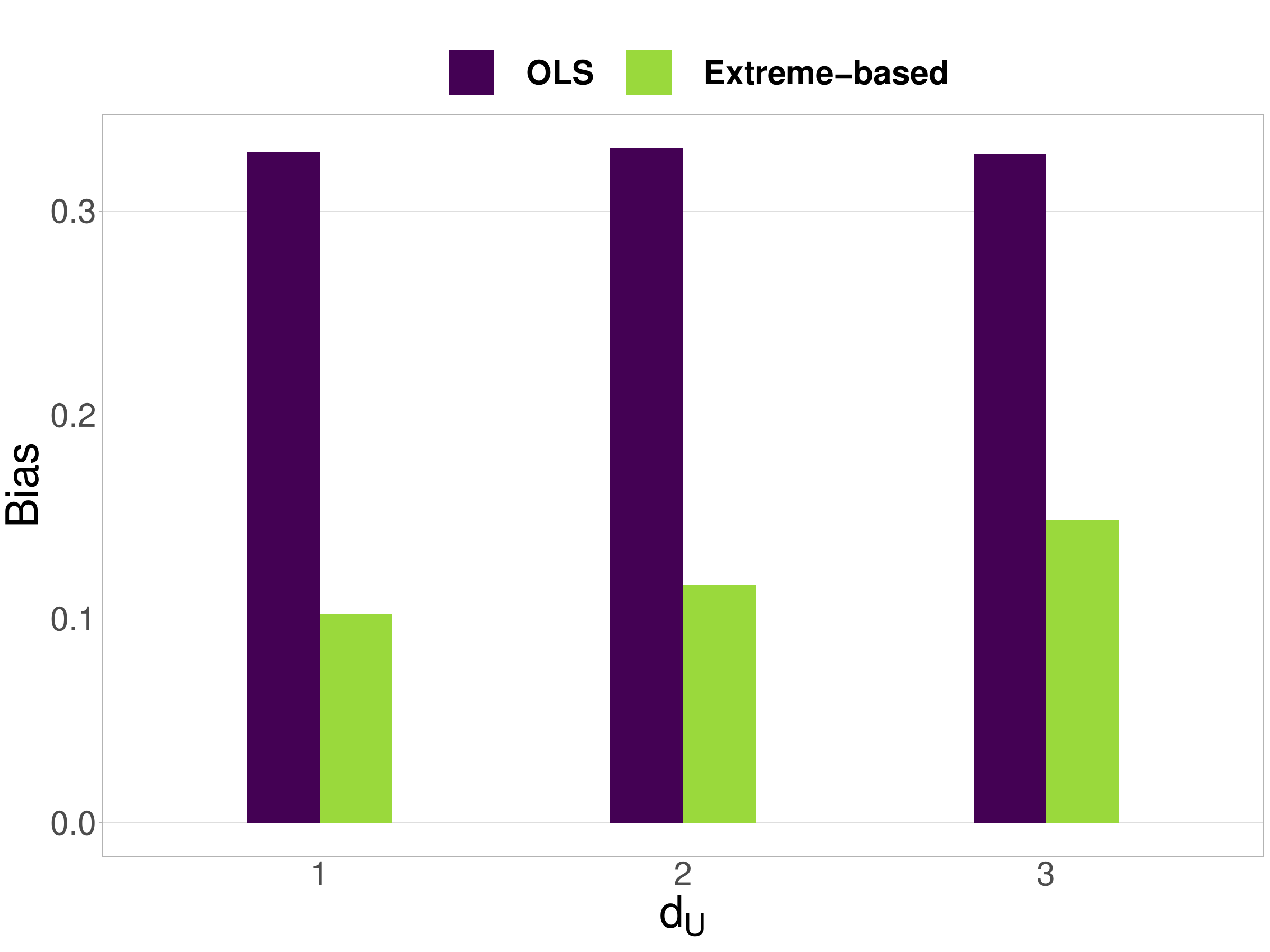}}
	\subfigure[MSE, $n = 5000$]{\includegraphics[scale = 0.18]{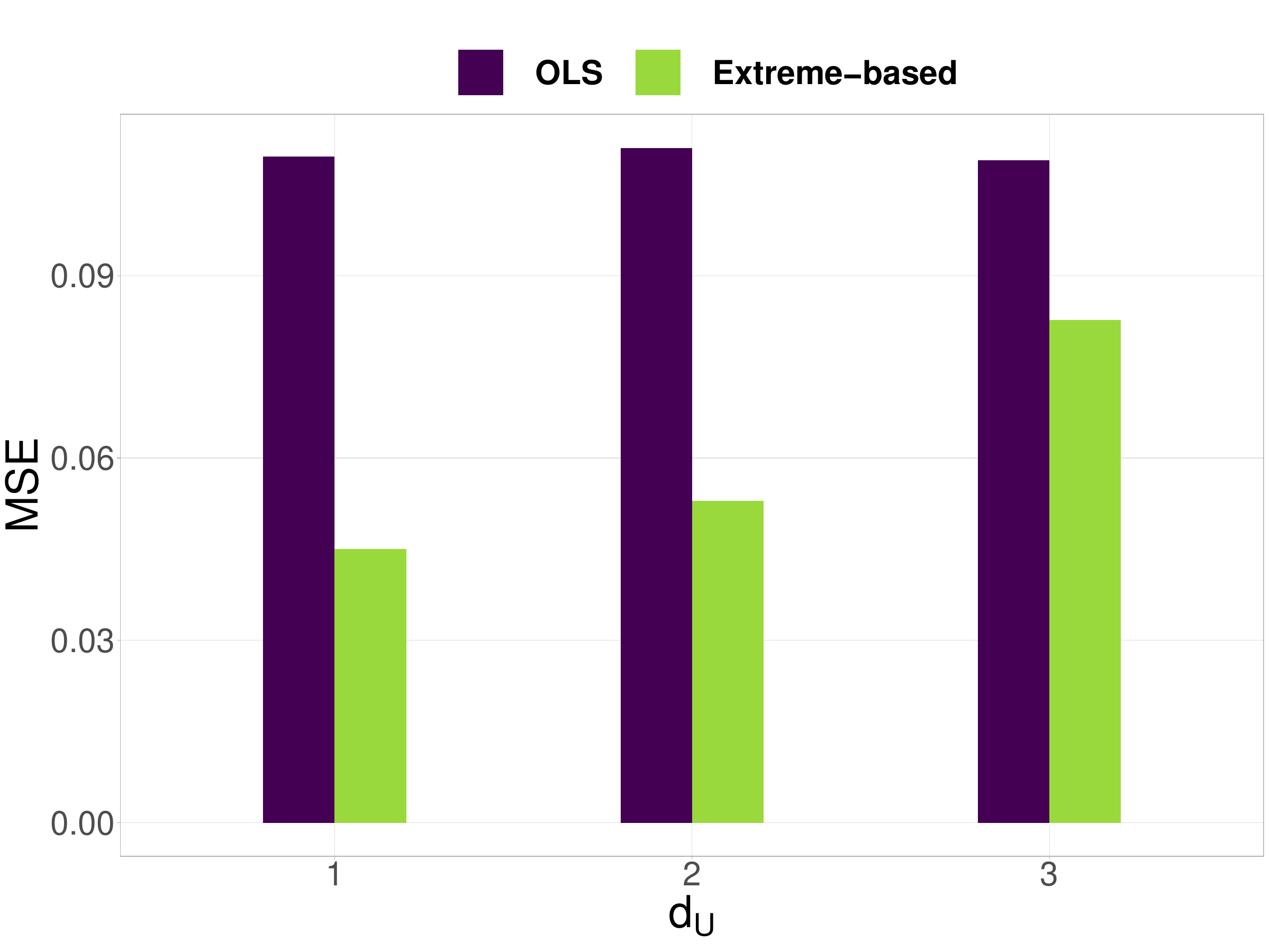}}
	\caption{Bias and MSE under the linear model with $\eta_{\epsilon} \sim  t(5)/2$ and different $d_{U}$ and $n$.}\label{fig: estimation heavy tail}
\end{figure}

We conduct an additional simulation study to investigate the performance of the extreme-based estimator with a heavy-tailed error term $\epsilon$. All the simulation settings are maintained except that we set $\eta_{\epsilon} \sim t(5)/2$ where $t(5)$ is the t-distribution with five degrees of freedom. Figure \ref{fig: estimation heavy tail} shows the biases and MSEs of the OLS estimator and the extreme-based estimator under different $d_{U}$ and $n$ with $\eta_{\epsilon} \sim t(5)/2$.

Comparing Fig.~\ref{fig: estimation heavy tail} with Fig.~\ref{fig: estimation}, it can be seen that the bias and MSE of the extreme-based estimator are larger when the error term is heavy-tailed, while those of the OLS estimator are similar under the two settings. The bias of both the OLS estimator and the extreme-based estimator does not decrease as $n$ increases when the error term is heavy-tailed. This implies both of these two estimators can not consistently estimate the causal effect $\theta_{0}$, which suggests the necessity of the light-tailedness of the error term in identifying the causal effect.

\section{Application to the Automobile Sale Dataset} \label{sec: automobile}
In this section, we apply our method to the automobile sale dataset from \cite{berry1995automobile} to investigate the causal effect of an automobile model's price on its utility. 
The dataset is accessible via the R package \emph{hdm}. 
The dataset includes $2217$ records on the price, market share, and various characteristics, such as the size and the ratio of horsepower to weight, of different automobile models marketed during the 20-year period from 1971 to 1990. 
The automobile industry is intensively studied in econometrics due to its large market size and economic importance \citep{berry1995automobile}. 
Researchers are particularly interested in the causal effect of the price on the sale of an automobile model or its utility for the customer.
However, statistical analysis of data from the automobile industry often suffers from endogeneity \citep{berry1995automobile}. 
In the following, we apply the proposed extreme-based method to mitigate this issue and estimate the causal effect of automobile price on the utility.

Following the analysis of \cite{berry1995automobile}, we treat each
model/year as an observation, as the characteristics of the same automobile model may vary across different years. For the $i$-th record with $i = 1,\dots, 2217$, let the exposure $X_{i}$ be the log price of the model in the corresponding year. 
Let  $S_{i}$ be the market share of the model in the corresponding year and $S_{0i}$ be the market share of the outside alternative in that year, that is, the market share of not purchasing any of the products in the dataset in that year. Under the model (6.3) in \cite{berry1995automobile}, the utility can be characterized by
\[
Y_{i} = \log(S_{i}) - \log(S_{0i}).
\]
In the following analysis, we employ a linear model 
\[
Y_{i} = \mu_{0} + X_{i}\theta_{0} + \epsilon_{i},
\]
where $\theta_{0}$ captures the causal effect of interest.

In order to estimate $\theta_{0}$, we conduct least squares regression and TSLS regression. In these regressions, we adjust for the ratio of horsepower to weight, a dummy for whether air conditioning is standard, miles per dollar and the size of the automobile model, which are the same observed characteristics of automobile models as those used in Section 7.3 of \cite{berry1995automobile}. To address serial correlation, we organize the records into $999$ disjoint clusters, aligning with the methodologies described in \cite{berry1995automobile} and \cite{andrews2017measuring}. Following \cite{berry1995automobile}, we treat observations across different clusters as independent and apply corrections for within-cluster correlation as outlined in their Section 5.2. The OLS estimator yields an estimate of $-0.082$ with $95\%$ confidence interval $[-0.086, -0.077]$. However, this estimate can be biased because both the error term and exposure are likely to be correlated with unobserved characteristics of the automobile model and its producer even after adjusting for the observed characteristics, which can lead to the problem of endogeneity.  Utilizing IVs from Berry’s study which consists of ten functions of the cost and demand characteristics of all products in a given year, the TSLS estimator produces an estimate of $-0.112$ with $95\%$ confidence interval $[-0.126 -0.099]$ after correcting for serials correlation. The confidence intervals produced by the OLS estimator and TSLS estimator do not overlap with each other. The TSLS estimator is considered more reliable due to its robustness against endogeneity.

In the dataset, $S_{0i}$ is always larger than $0.87$, which suggests that $S_{0i}$ is bounded away from zero. Note that $S_{i} \leq 1$. Thus, the outcome $Y_{i} = \log(S_{i}) - \log(S_{0i})$ is bounded from above. Then, it is reasonable to assume that the error is bounded from above and hence satisfies the light-tailedness condition (Assumption \ref{ass: tail prob}). Next, we implement the extreme-based estimator to estimate $\theta_{0}$.  We apply the extreme-based method at the cluster level to mitigate the problem of serial correlation. Suppose the outcome and exposure are centered and hence mean zero. For $m = 1,\dots, 999$, define the cluster-level exposure, outcome and error term of the $m$-th cluster as $\widetilde{X}_{m} = \sum_{r = 1}^{R_{m}}X_{m,r} / \sqrt{R_{m}}$, $\widetilde{Y}_{m} = \sum_{r = 1}^{R_{m}}Y_{m,r} / \sqrt{R_{m}}$ and $\widetilde{\epsilon}_{m} = \sum_{r = 1}^{R_{m}}\epsilon_{m,r} / \sqrt{R_{m}}$, respectively, where $R_{m}$ is the number of records in the $m$-th cluster, $X_{m,r}$, $Y_{m,r}$ and $\epsilon_{m,r}$ are the exposure, outcome and error term of the $r$-th record in the $m$-th cluster. The normalization constant $1/\sqrt{R_{m}}$ is adopted to ensure the variances of the cluster-level variables are similar across clusters with different sizes. Assuming that $\left\{\left(R_{m}, X_{m,1}, Y_{m,1}, \dots, X_{m,R_{m}}, Y_{m,R_{m}}\right)\right\}_{m = 1}^{999}$ is an i.i.d. sample, then $\{(\widetilde{X}_{m}, \widetilde{Y}_{m})\}_{m = 1}^{999}$ is an i.i.d. sample satisfying $\widetilde{Y}_{m} = \widetilde{X}_{m}\theta_{0} + \widetilde{\epsilon}_{m}$. We apply the extreme-based estimator with $\tau_{n} = 0.01/n^{1/4}$ to the cluster-level observations $\{(\widetilde{X}_{m}, \widetilde{Y}_{m})\}_{m = 1}^{999}$ to estimate the causal effect $\theta_{0}$. The resulting estimator produces an estimate of $-0.107$ with $95\%$ bootstrap confidence interval $[-0.120, -0.094]$, closely mirroring the TSLS result. This similarity suggests that our estimator can effectively address confounding issues without IVs.

\section{Discussion}
This paper proposes an extreme-based method, designed to effectively handle endogenous exposures and infer causal effects without relying on parametric assumptions or auxiliary variables. Central to the proposed extreme-based method is a light-tailedness condition on the error term, applicable to a variety of distributions, including the normal distribution. 

The proposed extreme-based method effectively addresses endogeneity under the light-tailed error condition by focusing on extreme observations. However, this focus may make the extreme-based estimator sensitive to outliers especially when $\tau$ is small. Thus, cautious interpretation of the results are required when there are potential outliers, and the data quality control is crucial for the success of the proposed extreme-based method. Despite these challenges, the proposed extreme-based method remains a valuable complement to the toolkit of causal effect estimation methods under endogeneity. The assumptions underlying various identification and estimation strategies all can be violated in practice. One advantage of having an additional strategy built on potentially plausible new assumptions is that it can reinforce causal conclusions when the results from different strategy are consistent. Moreover, the proposed extreme-based method is particularly reliable in scenarios where the outcome, $Y$, is derived from the average of raw data, effectively serving as a summary-level measure. For instance, in the real data example discussed in Section \ref{sec: automobile}, the market share is calculated as the average of numerous individual purchases. This averaging process mitigates the influence of any outliers present in the raw data, making the assumption of the light-tailed error more plausible, as supported by the central limit theorem.

The efficacy of the proposed extreme-based method depends crucially on the light-tailedness of the error distribution (Assumption \ref{ass: tail prob}). Nevertheless, variables, such as people's wealth, file sizes in computer systems and auction prices of art pieces, might be heavy-tailed in practice. It presents a valuable future direction to develop diagnostic tests for light-tailedness.

The proposed extreme-based method is specifically designed for structural equations with additive error terms. By applying a logarithmic transformation, multiplicative models can be reformulated as additive ones. However, this method is not universally applicable to general nonseparable structural equations. It is of interest to explore the possibility of identifying causal effects in general nonseparable models by utilizing the exposure--error relationship at the extremes.

	\bibliography{Additive}
	\bibliographystyle{plainnat}

	\newpage
	\setcounter{condition}{0}
	\setcounter{table}{0}
	\setcounter{algorithm}{0}
	\renewcommand{\thecondition}{\Alph{section}.\arabic{condition}}
	\renewcommand{\thetable}{\Alph{section}.\arabic{table}}
	\renewcommand{\thealgorithm}{\Alph{section}.\arabic{algorithm}}
	\appendix
	{\noindent \Huge \bf Appendix}
	\section{Regularity Conditions}\label{app: regularity cond}    
	The following conditions are the regularity conditions from \cite{chernozhukov2005extremal} required for the proof of Theorem \ref{thm: inference linear model}. Let $V = (1, X^{\T})^{\T}$. Subsequently, $a(t) \sim b(t)$ denotes $a(t) / b(t) \to 1$ as $t$ goes to some limit. 
	\begin{condition}\label{cond: R1}
		The distribution of $\epsilon$ is continuous and in the domain of attraction
		of generalized extreme value distributions with extreme value index $\omega$.
		Moreover, $P(\epsilon > t \mid X = x) \sim K(x)P(\epsilon > t)$ uniformly in $x \in \cX$ as $t \to q_{\epsilon}(0)$, where $K(\cdot)$ is a continuous bounded function.
	\end{condition}
	\begin{condition}\label{cond: R2}
		$\cX$ is compact and $E(VV^{\T})$ is positive definite.
	\end{condition}
	\begin{condition}\label{cond: R3}
		For $\omega$ in Condition \ref{cond: R1},
		(i) $\partial q_{\epsilon}(\tau, x)/\partial\tau \sim \partial q_{\epsilon}(\tau/ K(x))/\partial\tau$ uniformly in $x\in \cX$ as $\tau \to 0$; (ii) $\partial q_{\epsilon}(\tau)/\partial\tau$ is regularly varying at $0$ with exponent  $-\omega - 1$, i.e, 
		\[
		\lim_{\tau \to 0}\frac{\partial q_{\epsilon}(a\tau)/\partial\tau}{ \partial q_{\epsilon}(\tau)/\partial\tau} = a^{- \omega - 1}
		\]
		for every fixed $a > 0$.
	\end{condition}
	Conditions \ref{cond: R1}, \ref{cond: R2}, and \ref{cond: R3} are analogous to the Conditions R1, R2, and R3 of \cite{chernozhukov2005extremal}, respectively. Please refer to \cite{chernozhukov2005extremal} and the references therein for detailed explanations on the plausibility of these conditions. \cite{chernozhukov2005extremal} assumes that $q_{Y}(\tau, x)$ is linear in $x$, which is not adopted here. We invoke the following regularity condition instead.
	\begin{condition}\label{cond: limit var}
		The matrix of moments $E\{K(X)VV^{\T}\}$ exists and is nonsingular.
	\end{condition}
	Subsequently, we denote $E\{K(X)VV^{\T}\}$ by $\Sigma_{V}$.
	
	\section{Proofs}\label{app: proofs}
	In the proofs, we use $c$ and $C$ to denote generic positive constants that may differ in different places.
	\subsection{Proof of Example \ref{eg: unmeasured confounder}}
	\begin{proof}
		Under the latent unconfoundedness assumption and \eqref{eq: balance confounder}, we have
		\[
		\begin{aligned}
			\frac{p_{\epsilon\mid X}(e\mid x)}{p_{\epsilon}(e)} = \frac{\int p_{\epsilon\mid U}(e\mid u)p_{U\mid X}(u)du}{\int p_{\epsilon\mid U}(e\mid u)p_{U}(u)du} \geq \inf_{u}\left\{\frac{p_{U\mid X}(u\mid x)}{p_{U}(u)}\right\} \geq c_{x},
		\end{aligned}
		\]
		for some $c_{x} > 0$, $\forall e \in \cE$, and $x \in \cX$. The upper bound of the density ratio $p_{\epsilon\mid X}(e\mid x)/p_{\epsilon}(e)$ can be established similarly. These results verify Assumption \ref{ass: balance}. When $U\in\{1, \dots, K\}$ is a categorical variable and $p_{U\mid X}(u\mid x) > 0$ for any $u\in\{1,\dots, K\}$ and $x\in \cX$, we have $0 < \min_{k=1,\dots,K}p_{U}(k) \leq \max_{k=1,\dots,K}p_{U}(k) < 1$, $0 < \min_{k=1,\dots,K}p_{U\mid X}(k\mid x) \leq \max_{k=1,\dots,K}p_{U\mid X}(k\mid x) < 1$ and hence \eqref{eq: balance confounder} holds with $c_{x} = \min_{k=1,\dots,K}p_{U\mid X}(k\mid x) / \max_{k=1,\dots,K}p_{U}(k)$ and $C_{x} = \max_{k=1,\dots,K}p_{U\mid X}(k\mid x) / \min_{k=1,\dots,K}p_{U}(k)$.
		
		When $X$ is binary, we have $P(X = x)\geq c$ for $x = 0, 1$ under \eqref{eq: overlap}. Thus,
		\[
		c \leq \frac{P(X = x\mid U = u)}{P(X = x)} \leq c^{-1} - 1
		\]
		under \eqref{eq: overlap},
		which implies \eqref{eq: balance confounder} by noting that $p_{U\mid X}(u\mid x)/p_{U}(u) = P(X = x\mid U = u)/P(X = x)$. This completes the proof of Example \ref{eg: unmeasured confounder}.
	\end{proof}
	
	\subsection{Proof of Example \ref{eg: selection bias}}
	\begin{proof}
		It is straightforward to verify that, conditional on $S = 1$, $Y = f_{0}(X) + \epsilon$ with  $f_{0} = f_{*} + c^{*}$, $\epsilon = \epsilon^{*} - c^{*}$, and $c^{*} = E(\epsilon^{*}\mid S = 1)$. Then, $E(\epsilon\mid S = 1) = 0$ and
		\begin{equation}\label{eq: density ratio selection}
			\begin{aligned}
				\frac{p_{\epsilon\mid X, S}(e\mid x, 1)}{p_{\epsilon\mid S}(e\mid 1)}
				& = \frac{P(S = 1\mid X^{*} = x, \epsilon^{*} = e + c^{*})P(S = 1)p_{(X^{*},\epsilon^{*})}(x, e + c^{*})}{P(S = 1\mid \epsilon^{*} = e + c^{*})P(S = 1\mid X^{*} = x)p_{X^{*}}(x)p_{\epsilon^{*}}(e + c^{*})}\\
				& = \frac{P(S = 1\mid X^{*} = x, \epsilon^{*} = e + c^{*})P(S = 1)}{P(S = 1\mid \epsilon^{*} = e + c^{*})P(S = 1\mid X^{*} = x)}
			\end{aligned}
		\end{equation}
		for any $e \in \cE$, and $x \in \cX$,
		where the last equality holds because $X^{*}\Perp \epsilon^{*}$. Let $l_{*} = \inf_{x\in \cX,e \in \cE}P(S = 1\mid X^{*} = x, \epsilon^{*} = e + c^{*}) > 0$. Then, we have $P(S = 1), P(S = 1\mid \epsilon^{*} = e + c^{*}), P(S = 1\mid X^{*} = x) \in [l_{*}, 1]$. This together with \eqref{eq: density ratio selection} implies $c \leq p_{\epsilon^{*}\mid X^{*}, S}(e\mid x, 1)/p_{\epsilon^{*}\mid S}(e\mid 1) \leq C$.
		
		In addition, because $P(S = 1\mid X^{*}, Y^{*}) \geq c$, we have $p_{\epsilon^{*}\mid S}(e\mid 1)/p_{\epsilon^{*}}(e) = P(S = 1\mid \epsilon^{*} = e) / P(S = 1) = P(S = 1\mid Y^{*} - f(X^{*}) = e) / P(S = 1) \in [c, c^{-1}]$. Thus, Assumption \ref{ass: tail prob} is satisfied by the error $\epsilon$ conditional on $S = 1$ if it is satisfied by the original error $\epsilon^{*}$ in the absence of selection bias.
	\end{proof}
	
	\subsection{Proof of Example \ref{eg: measurement error}}
	\begin{proof}
		According to the independence assumption $U\Perp (W, V)$, the density of $(\epsilon, X)$ is
		\[
		p_{(\epsilon, X)}(e, x) = \int p_{V}(e + u^{\T}\theta_{0})p_{W}(x - u)p_{U}(u)du.
		\]
		Hence, we have
		\begin{equation}\label{eq: me dens error}
			p_{\epsilon}(e) = \int p_{V}(e + u^{\T}\theta_{0})p_{U}(u)du
		\end{equation}
		and
		\begin{equation}\label{eq: me cond dens error}
			p_{\epsilon\mid X}(e\mid x) = \int p_{V}(e + u^{\T}\theta_{0})p_{U}(u)\frac{p_{W}(x - u)}{\int p_{W}(x - t)p_{U}(t)dt}du.
		\end{equation}
		Then $c/C \leq p_{W}(x - u)/\int p_{W}(x - t)p_{U}(t)dt \leq C / c$ because $c \leq p_{W}(w) \leq C$. This implies $c/C \leq p_{\epsilon\mid X}(e\mid x) / p_{\epsilon}(e) \leq C / c$ according to \eqref{eq: me dens error} and \eqref{eq: me cond dens error}.
	\end{proof}
	
	\subsection{Proof of Proposition \ref{prop: quantile homo}}
	\begin{proof}
		If $\epsilon$ is upper bounded, then $q_{\epsilon}(\tau)$ and $q_{\epsilon}(x; \tau)$ converge to the upper bound of $\cE$ for any $x \in \cX$ as $\tau \to 0$ according to Assumption \ref{ass: balance}. The conclusion of the proposition holds in this case.
		
		If $\epsilon$ does not have an upper bound, then $P(\epsilon > t) > 0$ for any $t$ and $q_{\epsilon}(\tau) \to \infty$ as $\tau \to 0$.
		For any $\tau \in (0, 1)$, we have $P(\epsilon > q_{\epsilon}(\tau)) \leq \tau$ and $P(\epsilon > q_{\epsilon}(\tau)) \geq \tau$ according to the definition of the quantile. For any $\Delta > 0$ and $x \in \cX$, we have $P(\epsilon > q_{\epsilon}(\tau) - \Delta\mid X = x) \geq c_{x}P(\epsilon > q_{\epsilon}(\tau) - \Delta)$ for some constant $c_{x} > 0$ independent of $\tau$ by Assumption \ref{ass: balance}. Assumption \ref{ass: tail prob} implies that there is some function $\psi_{\Delta}(t)$ such that $\psi_{\Delta}(t) > 0$, $\lim_{t\to \infty}\psi_{\Delta}(t) = 0$, and 
		\[P(\epsilon > t ) \leq \psi_{\Delta}(t) P(\epsilon > t - \Delta ).\]
		Hence
		\[
		\begin{aligned}
			P(\epsilon > q_{\epsilon}(\tau) - \Delta\mid X = x)
			&\geq c_{x}P(\epsilon > q_{\epsilon}(\tau) - \Delta)\\
			&\geq c_{x}\psi_{\Delta}(q_{\epsilon}(\tau))^{-1}P(\epsilon > q_{\epsilon}(\tau))\\
			&> \tau,
		\end{aligned}
		\]
		for sufficiently small $\tau$, which implies
		\begin{equation}\label{eq: lower cq}
			q_{\epsilon}(x; \tau) \geq q_{\epsilon}(\tau) - \Delta.
		\end{equation}
		On the other hand, for any $\Delta > 0$ and $x \in \cX$,
		\[
		\begin{aligned}
			P(\epsilon > q_{\epsilon}(\tau) + \Delta\mid X = x)
			&\leq C_{x} P(\epsilon > q_{\epsilon}(\tau) + \Delta)\\
			&\leq C_{x}\psi_{\Delta}(q_{\epsilon}(\tau) + \Delta)P(\epsilon > q_{\epsilon}(\tau))\\
			&< \tau,
		\end{aligned}
		\]
		for any sufficiently small $\tau$ according to Assumptions \ref{ass: tail prob} and \ref{ass: balance}, and hence $q_{\epsilon}(x; \tau) \leq q_{\epsilon}(\tau) + \Delta$. This together with \eqref{eq: lower cq} and the arbitrariness of $\Delta$ implies that $|q_{\epsilon}(x; \tau) - q_{\epsilon}(\tau)| \to 0$ as $\tau \to 0$.
	\end{proof}
	
	\subsection{Proof of Theorem \ref{thm: identification}}
	Theorem \ref{thm: identification} is obtained from Proposition \ref{prop: quantile homo} and the arguments before Theorem \ref{thm: identification}. 
	
	\subsection{Proof of Lemma \ref{lemma: error bound}}   
	\begin{proof}
		Let $V = v(X)$ and $V_{i} = v(X_{i})$ for $i = 1,\dots,n$.
		For any $\beta$, according to Knight's identity \citep{knight1998limiting}, we have
		\begin{equation}\label{eq: knight equality}
			\begin{aligned}
				&\rho_{1 - \tau}(Y - V^{\T}\beta) - \rho_{1 - \tau}(Y - V^{\T}\barbeta)\\
				& = \rho_{1 - \tau}(Y - V^{\T}\barbeta - V^{\T}\delta) - \rho_{1 - \tau}(Y - V^{\T}\barbeta)\\
				& = - \left(1 - \tau - 1\{Y - V^{\T}\barbeta\leq 0\}\right)V^{\T}\delta + \int_{0}^{V^{\T}\delta}(1\{Y - V^{\T}\barbeta\leq s\} - 1\{Y - V^{\T}\barbeta\leq 0\})ds.
			\end{aligned}
		\end{equation}
		where $\delta = \beta - \barbeta$.
		Then
		\begin{equation}\label{eq: rough lower bound}
			\begin{aligned}
				&E\{\rho_{1 - \tau}(Y - V^{\T}\beta) - \rho_{1 - \tau}(Y - V^{\T}\barbeta)\mid X\}\\
				&= - E\left[1\{Y - q_{Y}(V; \tau) \leq 0\} - 1\{Y - V^{\T}\barbeta\leq 0\}\mid X\right]V^{\T}\delta\\
				&\quad + \int_{0}^{V^{\T}\delta}\left\{F_{Y\mid X}(V^{\T}\barbeta + s\mid X) - F_{Y\mid X}(V^{\T}\barbeta\mid X)\right\}ds\\
				&\geq - C_{f}\zeta_{a}|V^{\T}\delta| + \frac{f(V^{\T}\barbeta)}{2}|V^{\T}\delta|^{2} - \frac{1}{6}C_{L}|V^{\T}\delta|^{3},
			\end{aligned}
		\end{equation}
		where $F_{Y\mid X}$ is the distribution function of $Y$ conditional on $X$.
		According to Conditions \ref{cond: Lip} and \ref{cond: bound harzard}, we have
		\begin{equation}\label{eq: bound dens}
			\begin{aligned}
				f_{Y\mid X}(V^{\T}\barbeta\mid X) 
				&\geq f_{Y\mid X}(q_{Y}(X; \tau)\mid X)  - C_{L}\zeta_{a}\\
				&\geq c_{\tau} - C_{L}\zeta_{a}.
			\end{aligned}
		\end{equation}
		Note that $E(|V^{\T}\delta|^{3}) \leq \kappa_{3} \|\delta\|_{\Sigma}^{3}$ and $E(|V^{\T}\delta|) \leq \{E(|V^{\T}\delta|^{2})\}^{\frac{1}{2}} = \|\delta\|_{\Sigma}$. Combing this with \eqref{eq: rough lower bound} and \eqref{eq: bound dens}, we have
		\begin{equation}\label{eq: quadratic lower bound}
			\begin{aligned}
				E\{\rho_{1 - \tau}(Y - V^{\T}\beta) - \rho_{1 - \tau}(Y - V^{\T}\barbeta)\} &\geq
				\frac{1}{2}(c_{\tau} - C_{L}\zeta_{a})\|\delta\|_{\Sigma}^{2} - C_{f}\zeta_{a}\|\delta\|_{\Sigma} - \frac{1}{6}C_{L}\kappa_{3}\|\delta\|_{\Sigma}^{3}\\
				& \geq \frac{1}{6}(c_{\tau} - C_{L}\zeta_{a})\|\delta\|_{\Sigma}^{2},
			\end{aligned}
		\end{equation}
		if 
		\[
		\frac{6C_{f}\zeta_{a}}{c_{\tau} - C_{L}\zeta_{a}} \leq \|\delta\|_{\Sigma} \leq \frac{c_{\tau} - C_{L}\zeta_{a}}{C_{L}\kappa_{3}}.
		\]
		According to \eqref{eq: knight equality}, we have
		\begin{equation}\label{eq: bound infty}
			\begin{aligned}
				&|\rho_{1 - \tau}(Y - V^{\T}\beta) - \rho_{1 - \tau}(Y - V^{\T}\barbeta)| \leq \|V^{\T}\delta\|_{\infty} \leq \kappa_{\infty} \|\delta\|_{\Sigma},\\
				&\left|\frac{1}{n}\sum_{i=1}^{n}\rho_{1 - \tau}(Y_{i} - V_{i}^{\T}\beta) - \frac{1}{n}\sum_{i=1}^{n}\rho_{1 - \tau}(Y_{i} - V_{i}^{\T}\barbeta)\right| \leq \kappa_{\infty} \|\delta\|_{\Sigma},
			\end{aligned}
		\end{equation}
		and
		\[
		\begin{aligned}
			&|\rho_{1 - \tau}(Y - V^{\T}\beta) - \rho_{1 - \tau}(Y - V^{\T}\barbeta)|^{2} \\ 
			&\leq 2(1 - \tau - 1\{Y - V^{\T}\barbeta \leq 0\})^{2}|V^{\T}\delta|^{2} + 2(1\{Y - V^{\T}\barbeta \leq V^{\T}\delta\}\\
			&\quad - 1\{Y - V^{\T}\barbeta \leq 0\})^{2}|V^{\T}\delta|^{2}.
		\end{aligned}
		\]
		Thus
		\begin{equation}\label{eq: bound variance}
			E\{|\rho_{1 - \tau}(Y - V^{\T}\beta) - \rho_{1 - \tau}(Y - V^{\T}\barbeta)|^{2}\} \leq 2\{(\tau + C_{f}\zeta_{a})\|\delta\|_{\Sigma}^{2} + \kappa_{3} C_{f}\|\delta\|_{\Sigma}^{3}\} \colonequals \sigma^{2}(\|\delta\|_{\Sigma}).
		\end{equation} 
		According to \eqref{eq: bound infty} and \eqref{eq: bound variance}, the Bernstein inequality combined with union bound and standard covering number results (see Equation (5.9) in \cite{wainwright2019high}) can show that
		\begin{equation}\label{eq: prob bound}
			\begin{aligned}
				&P\Bigg(\sup_{\beta:\|\beta - \barbeta\|_{\Sigma} = r}
				\Bigg|
				E\left[\rho_{1 - \tau}(Y - V^{\T}\beta) - \rho_{1 - \tau}(Y - V^{\T}\barbeta)\right]\\
				&\phantom{sup_{\beta:\|\beta - \barbeta\| = r}} - \Bigg\{\frac{1}{n}\sum_{i=1}^{n}\rho_{1 - \tau}(Y_{i} - V_{i}^{\T}\beta) - \frac{1}{n}\sum_{i=1}^{n}\rho_{1 - \tau}(Y_{i} - V_{i}^{\T}\barbeta)\Bigg\}
				\Bigg| > 2t\Bigg) \\
				&\leq \exp\left\{-\frac{nt^{2}}{\sigma^{2}(r)} + p\log\left(1 + \frac{2r}{t}\right)\right\}
			\end{aligned}
		\end{equation}
		for any $r, t > 0$ provided $t \leq \sigma^{2}(r)/(\kappa_{\infty} r)$.
		For any $u > 0$, let 
		\[
		r = \frac{18}{c_{\tau} - C_{L}\zeta_{a}}\max\left\{\sqrt{\frac{(\tau + C_{f}\zeta_{a})(u + p\nu_{n,p,\tau})}{n}}, \frac{C_{f}\zeta_{a}}{3}\right\}
		\]
		and
		\[
		t = 2\sqrt{\frac{2(\tau + C_{f}\zeta_{a})(u + p\nu_{n,p,\tau})}{n}}r. 
		\]
		Under the conditions of Lemma \ref{lemma: error bound}, we have
		\[
		\frac{6C_{f}\zeta_{a}}{c_{\tau} - C_{L}\zeta_{a}} \leq r \leq \frac{c_{\tau} - C_{L}\zeta_{a}}{C_{L}\kappa_{3}},
		\]
		$t \leq \sigma^{2}(r)/(\kappa_{\infty} r)$,
		\[
		t < \frac{1}{6}(c_\tau - C_{L}\zeta_{a})r^{2},
		\]
		and
		\[
		\exp\left\{-\frac{nt^{2}}{\sigma^{2}(r)} + p\log\left(1 + \frac{2r}{t}\right)\right\} \leq \exp(- u).
		\]
		Then, combing \eqref{eq: quadratic lower bound} and \eqref{eq: prob bound}, we have 
		\[
		\inf_{\beta:\|\beta - \barbeta\| = r}\frac{1}{n}\sum_{i=1}^{n}\rho_{1 - \tau}(Y_{i} - V_{i}^{\T}\beta) > \frac{1}{n}\sum_{i=1}^{n}\rho_{1 - \tau}(Y_{i} - V_{i}^{\T}\barbeta)
		\]
		with probability at least $1 - \exp(-u)$.
		This implies
		\[
		\|\hbeta - \barbeta\|_{\Sigma} \leq r = \frac{18}{c_{\tau} - C_{L}\zeta_{a}}\max\left\{\sqrt{\frac{(\tau + C_{f}\zeta_{a})(u + p\nu_{n,p,\tau})}{n}}, \frac{C_{f}\zeta_{a}}{3}\right\}
		\]
		with probability at least $1 - \exp(-u)$ by the convexity of the check function, which proves the error bound of $\hbeta$.
		
		Note that 
		\[
		\begin{aligned}
			|q_{Y}(x;\tau) - q_{Y}(x_{0};\tau) - (v(x) - v(x_{0}))^{\T}\hbeta| 
			&\leq |q_{Y}(x;\tau) - q_{Y}(x_{0};\tau) - (v(x) - v(x_{0}))^{\T}\barbeta|\\
			&\quad + |(v(x) - v(x_{0}))^{\T}(\hbeta - \barbeta)|\\
			&\leq 2\zeta_{a} + 2\kappa_{\infty}\|\hbeta - \barbeta\|_{\Sigma}
		\end{aligned}
		\]
		and
		\[
		|\theta(x, x_{0}) - \{q_{Y}(x;\tau) - q_{Y}(x_{0};\tau)\}| \leq
		2\zeta_{\tau}(x, x_{0}).
		\]
		The error bound of $\hat{\theta}(x,x_{0})$ follows from the error bound of $\hbeta$.
	\end{proof}
	
	\subsection{Proof of Theorem \ref{thm: inference linear model}}\label{app: inference linear model}
	\begin{proof}
		Let $V_{i} = (1, X_{i}^{\T})^{\T}$ for $i = 1,\dots, n$ and
		\[
		R_{n} = \frac{\sqrt{n\tau_{n}}}{q_{\epsilon}(\tau_{n}) - q_{\epsilon}(\varpi\tau_{n})}
		\begin{pmatrix}
			\hat{\mu}_{n} - \mu_{0} - q_{\epsilon}(\tau_{n})\\
			\hat{\theta}_{n} - \theta_{0}
		\end{pmatrix}.
		\]
		According to Knight's identity \citep{knight1998limiting}, $R_{n}$ minimizes
		\[
		W_{n}(\tau_{n})^{\T}r + \Lambda_{n}(r, \tau_{n})
		\]
		with respect to $r$,
		where
		\[
		\begin{aligned}
			W_{n}(\tau_{n}) 
			& = -\frac{1}{\sqrt{n\tau_{n}}}\sum_{i=1}^{n}[1 - \tau_{n} - 1\{\epsilon_{i} \leq q_{\epsilon}(\tau_{n})\}]V_{i}\\
			& = \frac{1}{\sqrt{n\tau_{n}}}\sum_{i=1}^{n}[\tau_{n} - 1\{\epsilon_{i} > q_{\epsilon}(\tau_{n})\}]
			V_{i}
		\end{aligned}
		\]
		and for any $r = (r_{1}, r_{2}) \in \bbR \times \bbR^{d}$
		\[
		\begin{aligned}
			&\Lambda_{n}(r, \tau_{n})\\
			& = \frac{1}{q_{\epsilon}(\tau_{n}) - q_{\epsilon}(\varpi\tau_{n})}\sum_{i=1}^{n}\int_{0}^{\frac{(r_{1} + X_{i}^{\T}r_{2})}{\sqrt{n\tau_{n}}(q_{\epsilon}(\tau_{n}) - q_{\epsilon}(\varpi\tau_{n}))^{-1}}}\left[1\{\epsilon_{i}\leq q_{\epsilon}(\tau_{n}) + s\} - 1\{\epsilon_{i}\leq q_{\epsilon}(\tau_{n})\}\right]ds.
		\end{aligned}
		\]
		Note that
		\[
		\begin{aligned}
			E\{W_{n}(\tau_{n})\} 
			&= \sqrt{\frac{n}{\tau_{n}}}E([\tau_{n} - 1\{\epsilon > q_{\epsilon}(\tau_{n})\}]V)\\
			&= \sqrt{\frac{n}{\tau_{n}}}E([\tau_{n} - P\{\epsilon > q_{\epsilon}(\tau_{n})\mid X\}]V) \to 0
		\end{aligned}
		\]
		according to Condition \ref{cond: endogeneity bias} (i).
		Then, under Condition \ref{cond: R1}, we have
		\[
		\begin{aligned}
			\var\left\{W_{n}(\tau_{n})\right\} 
			&\sim E(\tau_{n}^{-1}P\{\epsilon > q_{\epsilon}(\tau_{n})\mid X\}VV^{\T}) 
			\\
			&\to E\{K(X)VV^{\T}\}\\
			& = \Sigma_{V}
		\end{aligned}
		\]
		as $n \to \infty$.
		According to Condition \ref{cond: limit var}, the central limit theorem implies
		\begin{equation}\label{eq: converge linear}
			W_{n}(\tau_{n}) - E\{W_{n}(\tau_{n})\} \to N(0, \Sigma_{V})
		\end{equation}
		in distribution as $n \to \infty$.
		In addition,
		by Condition \ref{cond: endogeneity bias} and Conditions \ref{cond: R1}--\ref{cond: limit var}, similar calculations as those in Equation (9.50) of \cite{chernozhukov2005extremal} can show that
		\[
		E\{\Lambda_{n}(r, \tau_{n})\} \to \frac{1}{2}r^{\T}Q_{\omega}r
		\] 
		as $n \to \infty$,
		where
		\[
		Q_{\omega} = \frac{2^{-\omega} - 1}{-\omega}E\left\{\frac{1}{K(X)^{\omega}}VV^{\T}\right\}
		\]
		if $\omega \neq 0$ and 
		\[
		Q_{\omega} = \log 2 E\left(VV^{\T}\right)
		\]
		if $\omega = 0$. Similarly to Lemma 9.6 (ii) of \cite{chernozhukov2005extremal}, we have $\var\{\Lambda_{n}(r, \tau_{n})\} \to 0$. Thus,
		\begin{equation*}\label{eq: converge quadratic}
			\Lambda_{n}(r, \tau_{n}) \to \frac{1}{2}r^{\T}Q_{\omega}r
		\end{equation*}
		in probability.  Because $R_{n}$ minimizes $W_{n}(\tau_{n})^{\T}r + \Lambda_{n}(r, \tau_{n})$ which is convex in $r$, we have
		\[
		R_{n} \to N(0, Q_{\omega}^{-1}\Sigma_{V}Q_{\omega}^{-1}),
		\]
		according to the convexity lemma \citep[see][p.826]{chernozhukov2005extremal}. This establishes the asymptotic normality with $\Sigma_{0}$ being the lower right $d\times d$ block of $Q_{\omega}^{-1}\Sigma_{V}Q_{\omega}^{-1}$.
	\end{proof}
	
	\subsection{Proof of Theorem \ref{thm: bootstrap}}
	The proof is similar to the second part of Theorem 2.2 in \cite{d2018extremal}. Let $R_{n}^{*}$, $W_{n}^{*}(\tau_{n})$, and $\Lambda_{n}^{*}(r, \tau_{n})$ be the bootstrap counterpart of $R_{n}$, $W_{n}(\tau_{n})$, and $\Lambda_{n}(r, \tau_{n})$ in the proof of Theorem \ref{thm: inference linear model}. Let $\{I_{n,j}\}_{j = 1}^{\infty}$ be an i.i.d. sequence from the multinomial distribution with size parameter $1$. number of events $n$, and probability $(1/n, \dots, 1/n)$. Define $p_{n, i} = \sum_{j = 1} 1\{I_{n, j } = i\}$. Then, according to Condition \ref{cond: endogeneity bias}, we have
	\begin{equation}\label{eq: boot linear}
		R_{n}^{*} = \frac{1}{\sqrt{n}}Q_{\omega}^{-1}\sum_{i=1}^{n}p_{n, i}\tau_{n}^{-1/2}(\tau_{n} - 1\{\epsilon > q_{\epsilon}(\tau_{n})\})
		V_{i} + o_{P}(1)
	\end{equation}
	by applying the same argument in the proof of Theorem 1
	in \cite{pollard1991asymptotics} and calculating the mean and variance of $W_{n}^{*}(\tau_{n})$, and $\Lambda_{n}^{*}(r, \tau_{n})$ similarly to the proof of Theorem \ref{thm: inference linear model}.
	
	The weights $\{p_{n, i}\}_{i = 1}^{n}$ are dependent.
	Next, we adopt the idea of Poissonization in Section 3.6.1 of \cite{vanweak} to remove the dependence.
	Let $N_{n}$ be a Poisson random variable with mean $n$, independent of the data and $\{I_{n,j}\}_{j = 1}^{\infty}$. Define $q_{n, i} = \sum_{j = 1}^{N_{n}}1\{I_{n, j } = i\}$ for $i = 1,\dots, n$. Then $\{q_{n, i}\}_{i = 1}^{n}$ are i.i.d. Poisson random variables with unit mean. Similarly to the proof of the second part of Theorem 2.2 in \cite{d2018extremal}, we have
	\[
	R_{n}^{*} - R_{n} = \frac{1}{\sqrt{n}}Q_{\omega}^{-1}\sum_{i=1}^{n}(q_{n, i} - 1)\tau_{n}^{-1/2}(\tau_{n} - 1\{\epsilon > q_{\epsilon}(\tau_{n})\})
	V_{i}  + o_{P}(1).
	\]
	Noting that $E(q_{n, i} - 1) = 0$ and $\var(q_{n, i} - 1) = 1$, we have
	\[
	R_{n}^{*} - R_{n} \to N(0, Q_{\omega}^{-1}\Sigma_{V}Q_{\omega}^{-1})
	\]
	in distribution conditional on the data with probability approaching one  as $n \to \infty$ according to Lemma 2.9.5 in \cite{vanweak}. This establishes the conclusion of Theorem \ref{thm: bootstrap}.
	
	\section{Implementation Details}\label{app: implement}
	\subsection{Select the Tail}\label{app: select tail}
	In practice, it is possible that a lower tail probability condition holds instead of Assumption \ref{ass: tail prob}.
	\begin{condition}\label{cond: lower tail prob}
		For any $\Delta > 0$, we have $P(\epsilon \leq -t ) = o(1)P(\epsilon \leq -t + \Delta )$ as $t \to \infty$.
	\end{condition} 
	Condition \ref{cond: lower tail prob} is a counterpart of Assumption \ref{ass: tail prob} on the lower tail of the error term. Under Assumption \ref{ass: balance} and Condition \ref{cond: lower tail prob}, the results of Proposition \ref{prop: quantile homo} and Theorem \ref{thm: identification} hold as $\tau \to 1$. In this case, $\theta(x, x_{0})$ can be estimated by 
	\[
	\hat{\theta}_{L}(x, x_{0}) = \{v(x) - v(x_{0})\}^{\T}\hbeta_{L},
	\] 
	where 
	\[
	\hbeta_{L} = \mathop{\arg\min}_{\beta} \frac{1}{n}\sum_{i=1}^{n}\rho_{\tau}(Y_{i} - v(X_{i})^{\T}\beta),
	\]
	
	In practice, it may be unclear whether Assumption \ref{ass: tail prob} or Condition \ref{cond: lower tail prob} is more applicable. Consequently, it becomes important to ascertain in a data-driven manner whether to utilize $\hat{\theta}(x, x_{0})$ or $\hat{\theta}_{L}(x, x_{0})$. To this end, we define the upper residuals $\hat{\epsilon}_{U,i} = Y_{i} - \hat{\theta}(X_{i}, x_{0})$ and lower residuals $\hat{\epsilon}_{L,i} = Y_{i} - \hat{\theta}_{L}(X_{i}, x_{0})$ for $i = 1,\dots, n$. If Assumption \ref{ass: tail prob} holds, we have $\hat{\epsilon}_{U,i} \approx Y_{i} - f_{0}(X_{i}) + f_{0}(x_{0})$. Then Proposition \ref{prop: quantile homo} implies that the upper extreme conditional quantiles of the upper residual are nearly independent of the exposure. Conversely, under Condition \ref{cond: lower tail prob}, the lower extreme conditional quantiles of the lower residual should display independence from $X$. Thus, we take a set of grid points $x_{1}, \dots, x_{K}$ in $\cX$ and estimate the conditional extreme quantiles of the residuals at $x_{k}$ for $k = 1,\dots, K$. Let $\cI_{k}$ be the index set of the observations whose exposure value is among the $n/\log(n)$ nearest to $x_{k}$. Define
	$\hat{q}_{U,k} = \max_{i \in \cI_{k}}\hat{\epsilon}_{U,i}$ and $\hat{q}_{L,k} = \min_{i \in \cI_{k}}\hat{\epsilon}_{L,i}$. We adopt $\hat{\theta}(x, x_{0})$ as the final estimator for the causal effect $\theta(x, x_{0})$ when the range of $\hat{q}_{U,k}$ is smaller than that of $\hat{q}_{L,k}$; specifically, if $\max_{k}\hat{q}_{U,k} - \min_{k}\hat{q}_{U,k} < \max_{k}\hat{q}_{L,k} - \min_{k}\hat{q}_{L,k}$. Otherwise, we use $\hat{\theta}_{L}(x, x_{0})$.
	
	\subsection{Adjust for Covariates}\label{app: adjust for covariates}
	Suppose $X$ contains both exposures of interest and covariates. Let $\cA$ and $\cC$ be the index sets of the exposures and covariates. For any vector $a$, let $a_{\cA}$ and $a_{\cC}$ be the subvectors of $a$ consist of components in $\cA$ and $\cC$, respectively. Then, $X = (X_{\cA}^{\T}, X_{\cC}^{\T})^{\T}$. Suppose for given values $x_{\cA}$ and $x_{\cA, 0}$ of the exposure, the parameter of interest is the average causal effect $\theta_{\cA}(x_{\cA}, x_{\cA, 0}) = E\{f(x_{\cA}, X_{\cC})\} - E\{f(x_{\cA, 0}, X_{\cC})\}$. Let $v(x_{\cA}, x_{\cC})=\{v_{j}(x_{\cA}, x_{\cC}): j = 1,\dots, p\}^\T$ be a set of basis functions with $v_{1}(x_{\cA}, x_{\cC}) \equiv 1$. Then, the average causal effect $\theta_{\cA}(x_{\cA}, x_{\cA, 0})$ can be estimator by the modified extreme-based estimator
	\[
	\hat{\theta}_{\cA}(x_{\cA}, x_{\cA, 0}) = \frac{1}{n}\sum_{i= 1}^{n}\{v(x_{\cA}, X_{i, \cC}) - v(x_{\cA, 0}, X_{i, \cC})\}^{\T}\hbeta.
	\]
	where 
	$\hbeta = \mathop{\arg\min}_{\beta} n^{-1}\sum_{i=1}^{n}\rho_{1 - \tau}(Y_{i} - v(X_{i,\cA}, X_{i, \cC})^{\T}\beta)$
	and $\rho_{1 - \tau}(z) = z(1 - \tau-1\{z < 0\})$.
	
	There is an alternative way to construct extreme-based causal effect estimator when the relationship between $Y$ and $X$ is linear and the covariate $X_{\cC}$ is known to be exogenous. Specifically, suppose 
	\[
	Y = \mu_{0} + X_{\cA}^{\T}\theta_{\cA, 0} + X_{\cC}^{\T}\theta_{\cC, 0} + \epsilon,
	\]
	where $X_{\cC} \Perp \epsilon$.
	Define $(\mu_{Y}, \gamma_{Y}) = \mathop{\arg\min}_{\mu, \gamma}E\{(Y - \mu - X_{\cC}^{\T}\gamma)^{2}\}$ and $(\mu_{\cA}, \Gamma_{\cA}) = \mathop{\arg\min}_{\mu, \Gamma}E\{\|X_{\cA} - \mu - \Gamma^{\T} X_{\cC}\|^{2}\}$ where $\|\cdot\|$ denotes the Euclid norm. Let $\xi_{Y} = Y - \mu_{Y} - X_{\cC}^{\T}\gamma_{Y}$ and $\xi_{\cA} = X_{\cA} - \mu_{\cA} - \Gamma_{\cA}^{\T} X_{\cC}$. Then, we have
	\[
	\xi_{Y} = \xi_{\cA}^{\T}\theta_{\cA, 0} + \epsilon.
	\] 
	This relationship can be used to estimate the causal effect $\theta_{\cA, 0}$.
	For $i = 1,\dots, n$, let
	\[
	\hat{\xi}_{Y, i} = Y_{i} - \hat{\mu}_{Y} - X_{i, \cC}^{\T}\hat{\gamma}_{Y}
	\]
	and
	\[
	\hat{\xi}_{\cA, i} = X_{i, \cA} - \hat{\mu}_{\cA} - \widehat{\Gamma}_{\cA}^{\T} X_{i, \cC},
	\]
	where $\hat{\mu}_{Y}$, $\hat{\gamma}_{Y}$, $\hat{\mu}_{\cA}$, and $\widehat{\Gamma}_{\cA}$ are the sample counterparts of  $\mu_{Y}$, $\gamma_{Y}$, $\mu_{\cA}$, and $\Gamma_{\cA}$, respectively.
	Then, $\theta_{\cA, 0}$ can be estimated by 
	\[
	\hat{\theta}_{\cA, n} = \mathop{\arg\min}_{\theta} \frac{1}{n}\sum_{i=1}^{n}\rho_{1 - \tau_{n}}(\hat{\xi}_{Y, i} - \xi_{\cA, i}^{\T}\theta).
	\]
	We propose the theoretical analysis of the estimators $\hat{\theta}_{\cA}(x_{\cA}, x_{\cA, 0})$ and $\hat{\theta}_{\cA, n}$ as a topic for future research.
	
	\subsection{Selection of the Tail Index $\tau$}\label{app: select tau}
	The proposed extreme-based method uncovers the causal effect by exploiting the information at the extreme quantiles, which involves the tail index $\tau$ as a ``tuning parameter". Theoretical results in the main text provide some guideline for selecting the tail index $\tau$. In this section, we propose a data-adaptive procedure for selecting $\tau$ which might be useful in practical implementation.
	
	Intuitively, one faces the bias-variance trade-off when selecting $\tau$. When $\tau$ is small, the bias caused by endogeneity tends to be small according to Theorem \ref{thm: identification}, while the variance of the extreme-based estimator tends to be large because only a small fraction of observations is informative in estimating the upper $\tau$-quantile in this case. By increasing $\tau$, the variance can be reduced at the cost of possible bias increase. We approximate the bias and variance utilizing a bootstrap procedure and select $\tau$ based on the approximated bias and variance.
	
	In this section, we use $\hat{\theta}_{\tau}(x, x_{0})$ to denote the extreme-based estimator defined using the tail index $\tau$. Suppose $\cT = \{\tau_{m}\}_{m=1}^{M}$ is a candidate set for $\tau$ and $B$ is a user-specified large integer. We select $\tau$ using the following algorithm. 
	
	\begin{breakablealgorithm}
		\caption{The algorithm for data-adaptive selection of $\tau$}
		\begin{algorithmic}[1]
			\State {\bf Input}: $\{(X_{i}, Y_{i})\}_{i = 1}^{n}$, $\cT$ and $B$.
			\For{$\tau \in \cT$}
			\For{$b = 1,\dots, B$}
			\State Draw a sample $\left\{(X_{i}^{(b)}, Y_{i}^{(b)})\right\}_{i = 1}^{n}$ with replacement from $\left\{(X_{i}, Y_{i})\right\}_{i = 1}^{n}$;
			\State Calculate
			\[
			\hbeta_{\tau}^{(b)} = \mathop{\arg\min}_{\beta} \frac{1}{n}\sum_{i=1}^{n}\rho_{1 - \tau}(Y_{i} - v(X_{i})^{\T}\beta)
			\]
			and
			\[
			\hbeta_{\tau/2}^{(b)} = \mathop{\arg\min}_{\beta} \frac{1}{n}\sum_{i=1}^{n}\rho_{1 - \tau/2}(Y_{i} - v(X_{i})^{\T}\beta);
			\]
			\State Calculate 
			\[
			\hat{\theta}_{\tau}^{(b)}(x, x_{0}) = \{v(x) - v(x_{0})\}^{\T}\hbeta_{\tau}^{(b)}
			\] and 
			\[
			\hat{\theta}_{\tau/2}^{(b)}(x, x_{0}) = = \{v(x) - v(x_{0})\}^{\T}\hbeta_{\tau/2}^{(b)};
			\]
			\EndFor
			\State Calculate 
			\[
			\widehat{\rm bias}(\tau) = \frac{1}{B} \sum_{b = 1}^{B}\hat{\theta}_{\tau}^{(b)}(x, x_{0}) - \frac{1}{B} \sum_{b = 1}^{B}\hat{\theta}_{\tau/2}^{(b)}(x, x_{0}),
			\]
			and
			\[
			\widehat{\rm var}(\tau) = \frac{1}{B}\sum_{b = 1}^{B}\left\{\hat{\theta}_{\tau}^{(b)}(x, x_{0}) - \frac{1}{B}\sum_{b = 1}^{B}\hat{\theta}_{\tau}^{(b)}(x, x_{0})\right\}^{2};
			\]
			\EndFor 
			\State Select $\hat{\tau}_{s} = \mathop{\arg\min}_{\tau\in\cT}\left\{
			\widehat{\rm bias}(\tau)^2 + \widehat{\rm var}(\tau)\right\}$;
			\State {\bf Output:}  $\hat{\tau}_{s}$.
		\end{algorithmic}
		\label{alg: sel tau}
	\end{breakablealgorithm}
	According to Theorem \ref{thm: identification}, the bias of the extreme-based estimator tends to be small when $\tau$ is small. For any $\tau \in \cT$, we use the mean $\sum_{b = 1}^{B}\hat{\theta}_{\tau/2}^{(b)}(x, x_{0}) / B$ of the bootstrapped extreme-based estimator with tail index $\tau / 2$ to approximate the true causal effect and estimate $\hat{\theta}_{\tau}(x, x_{0})$'s bias by $\widehat{\rm bias}(\tau)$ in Algorithm \ref{alg: sel tau}. Moreover, $\widehat{\rm var}(\tau)$ is the bootstrap estimation of $\hat{\theta}_{\tau}(x, x_{0})$'s variance.  Utilizing $\widehat{\rm bias}(\tau)$ and $\widehat{\rm var}(\tau)$, Algorithm \ref{alg: sel tau} selects the index $\hat{\tau}_{s}$ that minimizes the approximated MSE $\widehat{\rm bias}(\tau)^2 + \widehat{\rm var}(\tau)$ over $\cT$.
	
	Table \ref{table: sel tau} presents the MSE of the extreme-based estimator with non-adaptive and data-adaptive $\tau$ under the simulation settings in Section \ref{sec: sim} in the main text. For the non-adaptive extreme-based estimator, we set $\tau = 0.01 / n^{1/4}$. For the data-adaptive extreme-based estimator, $\tau$ is selected from the candidate set $\cT = \{0.01 \times k / n^{1/4}: k = 1,\dots, 5\}$ using Algorithm \ref{alg: sel tau}. The results in Table \ref{table: sel tau} demonstrate that employing a data-adaptive $\tau$ consistently improves the MSE of the extreme-based estimator across various settings with different combinations of $n$ and $d_{U}$.

	\begin{table}
		\caption{MSE of the extreme-based estimator with non-adaptive and data-adaptive tail index}\label{table: sel tau} 
		\begin{tabular}{*{7}{c}}
			\toprule
			$(n, d_{U})$ & $(1000, 1)$ & $(1000, 2)$ & $(1000, 3)$ & $(5000, 1)$ & $(5000, 2)$ & $(5000, 3)$\\
			\midrule
			non-adaptive & $0.0108$ & $0.0124$ & $0.0166$ & $0.0039$ & $0.0053$ & $0.0047$ \\
			data-adaptive & $0.0094$ & $0.0110$ & $0.0140$ & $0.0035$ & $0.0038$ & $0.0041$ \\
			\bottomrule
		\end{tabular}
	\end{table}
	
	\section{Repair Invalid Auxiliary Variables}\label{app: inference invalid auxiliary}
	In this section, we apply proposed extreme-based approach to the invalid auxiliary variable problem. Assume $X$ is scalar and the outcome $Y$ satisfies
	\[
	Y = \mu_{0} + X\theta_{0} + \epsilon,
	\]
	where $\theta_{0}$ is the causal effect of interest and $\epsilon$ is correlated with $X$. Researchers may resort to auxiliary variables that meet certain assumptions, such as IVs or negative controls, to address the endogeneity problem. However, as discussed in  Section \ref{sec: identification}, it is often challenging to verify the validity of auxiliary variables in practice, which leads to the invalid auxiliary variable problem. Let $Z$ be a $d_{z}$-dimensional vector of candidate auxiliary variables where the dimension $d_{z}$ is fixed. For $j = 1,\dots, d_{z}$, let $Z_{j}$ and $Z_{-j}$ be the $j$-th component of $Z$ and the subvector of $Z$ that excludes the $j$-th component. 
	\begin{remark}
		For the IV approach, $Z$ is a vector of candidate IVs. For the double negative control approach \citep{miao2018identifying,cui2023semiparametric}, to be specific, we suppose that $Z$ is a vector of candidate negative control exposures and assume in addition that a valid negative control outcome is available.
	\end{remark}
	For $\alpha \in (0, 1)$, suppose the auxiliary variable-based estimator  that treats $Z_{j}$ as the auxiliary variable and $Z_{-j}$ as covariates
	can produce an asymptotic $(1-\alpha)$-confidence interval $[\hat{\theta}_{L,j}(\alpha), \hat{\theta}_{U,j}(\alpha)]$ for $\theta_{0}$ if $Z_{j}$ is a valid auxiliary variable.
	This statement is true for many standard auxiliary variable approaches such as the two-stage least squares (TSLS) method \citep{sargan1958estimation} for IV estimation and the confounding bridge method \citep{miao2018identifying} for double negative control estimation. In practice, we don't know whether or not $Z_{j}$ is a valid auxiliary variable. On the other hand, by leveraging the light-tailedness of the error term, 
	the causal effect can be consistently estimated by the proposed extreme-based estimator $\hat{\theta}_{n}$ without utilizing IVs. 
	However, the convergence rate of $\hat{\theta}_{n}$ may be slower than $1/\sqrt{n}$ according to the discussion behind Theorem \ref{thm: inference linear model}. In contrast, the auxiliary variable-based estimator, such as the TSLS estimator for IV estimation or the confounding bridge estimator for double negative control estimation, can be $\sqrt{n}$-consistent under regularity conditions when valid auxiliary variables are used \citep{hayashi2011econometrics}. Then, the lengths of the valid auxiliary variable-based confidence intervals is of order $1/\sqrt{n}$ which can be shorter in order than the extreme-based confidence intervals. This motivates us to select valid IVs based on the consistent estimator $\hat{\theta}_{n}$ and construct the confidence set based on the selected auxiliary variables.

	Let $\cV$ be the index set of valid auxiliary variables. Next, we select valid IVs based on $\hat{\theta}_{n}$. Recall that $\hat{c}_{\alpha}$ is the $(1 - \alpha)$-bootstrap quantile defined in the last section for any $\alpha \in (0, 1)$. If $Z_{j}$ is a valid auxiliary variable, then both $[\hat{\theta}_{L,j}(\alpha), \hat{\theta}_{U,j}(\alpha)]$ and $[\hat{\theta}_{n} - \hat{c}_{\alpha}, \hat{\theta}_{n} + \hat{c}_{\alpha}]$ contains $\theta_{0}$ and hence overlap with each other with high probability. The valid auxiliary variable selection procedure is built on top of this observation. Specifically, for $\lambda \in (0, 1)$, let 
	\[\widehat{\cV} = \left\{j : [\hat{\theta}_{L,j}(\lambda\alpha / 2), \hat{\theta}_{U,j}(\lambda\alpha / 2)] \cap[\hat{\theta}_{n} - \hat{c}_{\lambda\alpha / 2}, \hat{\theta}_{n} + \hat{c}_{\lambda\alpha / 2}] \neq \emptyset\right\}\]
	be the estimated index set of valid auxiliary variables. Then, we define the $(1 - \alpha)$-confidence set for $\theta_{0}$ as
	\begin{equation}\label{eq: IV CI}
		\bigcup_{j \in \widehat{\cV}}[\hat{\theta}_{L,j}(\alpha - \lambda\alpha), \hat{\theta}_{U,j}(\alpha - \lambda\alpha)].
	\end{equation}
	The confidence interval \eqref{eq: IV CI} is asymptotically valid as long as at least one auxiliary variable is valid. To see this, suppose the $j_{\star}$-th auxiliary variable is valid. Then, asymptotically, $j_{\star} \in \widehat{\cV}$ with probability no less than $1 - \lambda\alpha$. Hence, asymptotically, $\theta_{0} \in [\hat{\theta}_{L,j_{\star}}(\alpha - \lambda\alpha), \hat{\theta}_{U,j_{\star}}(\alpha - \lambda\alpha)] \subset \bigcup_{j \in \widehat{\cV}}[\hat{\theta}_{L,j}(\alpha - \lambda\alpha), \hat{\theta}_{U,j}(\alpha - \lambda\alpha)]$ with probability no less than $1 - \lambda\alpha - (1 - \lambda)\alpha = 1 - \alpha$, which establish the validity of the confidence set \eqref{eq: IV CI}.
	Furthermore, suppose the lengths of the auxiliary variable-based confidence intervals is of order $\Theta(1/\sqrt{n})$.  Then, the lengths of the extreme-based confidence interval and the confidence set \eqref{eq: IV CI} are asymptotically of orders $\Theta(\{q_{\epsilon}(\tau_{n}) - q_{\epsilon}(\varpi \tau_{n})\}/\sqrt{n\tau_{n}})$ and $\Theta(1/\sqrt{n})$, respectively. Thus, the confidence set based on the selected auxiliary variables is expected to be shorter than that solely based on $\hat{\theta}_{n}$ when the sample size is large and $\{q_{\epsilon}(\tau_{n}) - q_{\epsilon}(\varpi \tau_{n})\}/\sqrt{n\tau_{n}} \gg 1 / \sqrt{n}$.
	
	
	The confidence set \eqref{eq: IV CI} is valid for any $\lambda \in (0, 1)$. When $\lambda$ is small, each interval in the union \eqref{eq: IV CI} tends to be short, but the set $\widehat{\cV}$ tends to be large. On the other hand, as $\lambda$ increases, each interval in the union \eqref{eq: IV CI} becomes longer, and the set $\widehat{\cV}$ becomes smaller. In practice, one can try multiple $\lambda$'s in a finite candidate set such as $\{0.05, \dots, 0.95\}$ and choose the $\lambda$ that minimizes the length of the resulting confidence set. Let $\widehat{\lambda}$ be the resulting $\lambda$. The coverage probability of the resulting confidence set can be guaranteed asymptotically provided $P(\widehat{\lambda} = \lambda_{*}) \to 1$ for some $\lambda_{*}$ in the candidate set for $\lambda$.
	
	To construct a valid confidence set for the causal effect, existing methods that can accommodate invalid auxiliary variables usually focus on the invalid IV problem.
	In contrast, our procedure applies to general causal inference problem with invalid auxiliary variables.
	In the context of invalid IV, existing methods often impose restrictions on the number of valid IVs \citep{kang2016instrumental, guo2018confidence, windmeijer2021confidence, lin2024instrumental} or assumptions on the form of the IVs' effects on the outcome and exposure \citep{tchetgen2021genius, sun2023semiparametric,ye2024genius, guo2024robustness}. Our procedure does not rely on such assumptions and can effectively operate even with a single unknown valid IV.
	
	\section{Simulations with Possibly Invalid IVs}\label{app: sim invalid IV}
	We consider the scenario where some possibly invalid IVs $Z$ are available in addition to $Y$ and $X$. The unmeasured confounder $U$ is generated in the same way as in Section \ref{sec: sim}. Suppose $Z_{1}, Z_{2}, Z_{3} \Perp U$ are the candidate IVs, where $Z_{1}$, $Z_{2}$, and $Z_{3}$ are independent and follow ${\rm Bernoulli(0.5)}$. The exposure $X$ and outcome $Y$ follow the linear models
	\[
	X = 2Z_{1} + 2Z_{2} + 2 Z_{3} + 0.2U^{\T}\gamma_{U} + 
	\eta_{X}
	\]  
	and
	\[
	Y = 2Z_{2} + 2Z_{3} + X\theta_{0} + 4U^{\T}\gamma_{U} 
	+ \eta_{\epsilon},
	\]
	respectively. The parameters and error terms $\theta_{0}$, $\gamma_{U}$, $\eta_{X}$ and $\eta_{\epsilon}$ are set in the same way as in Section \ref{sec: sim}. Under this simulation setting, $Z_{1}$ is a valid IV while $Z_{2}$ and $Z_{3}$ are invalid IVs. Neither the majority valid nor the plurality valid assumption holds in this setting. 
	
\begin{figure}[h]
		\centering
		\subfigure[Bias, $n = 1000$]{\includegraphics[scale = 0.18]{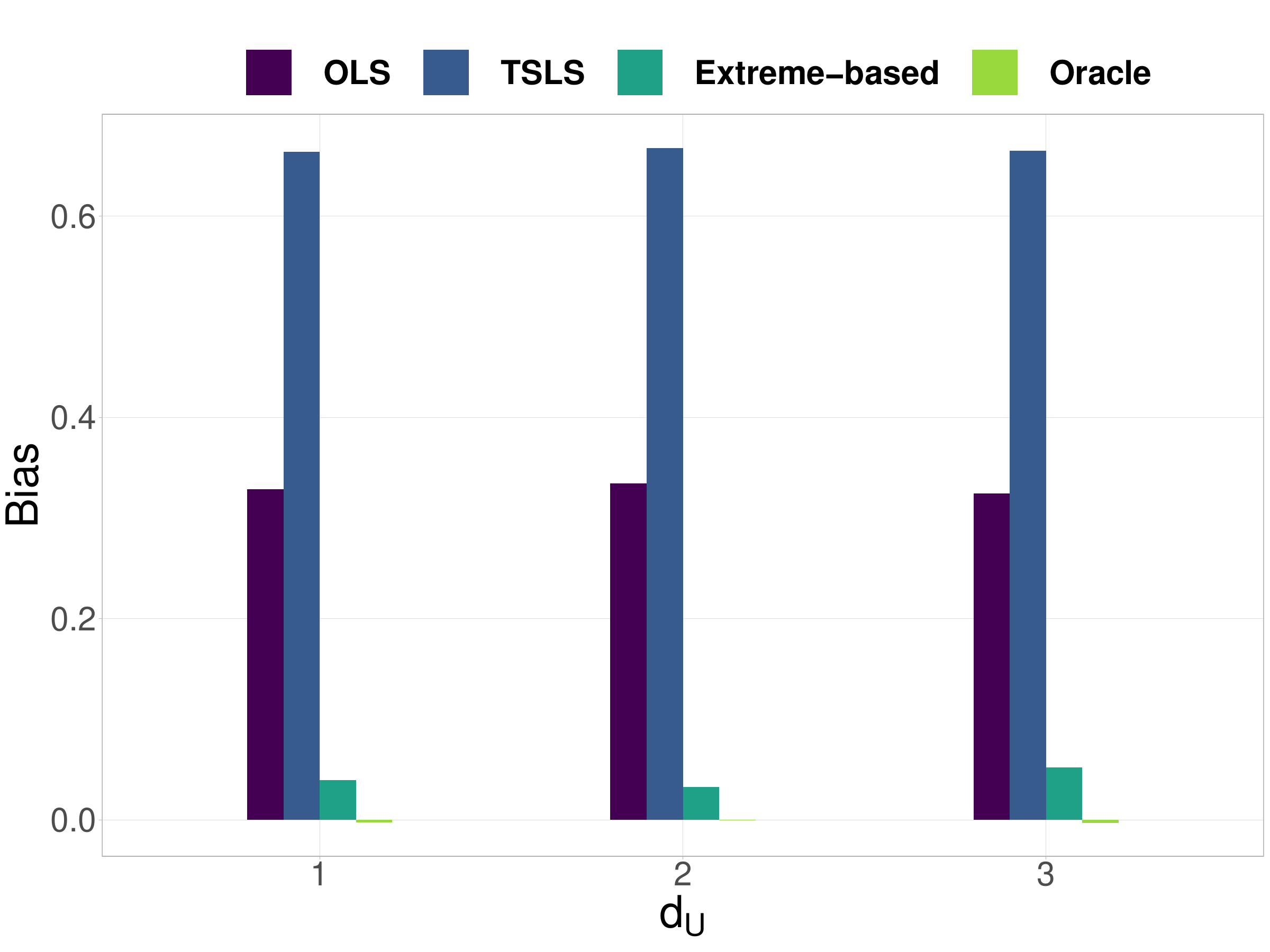}}
		\subfigure[MSE, $n = 1000$]{\includegraphics[scale = 0.18]{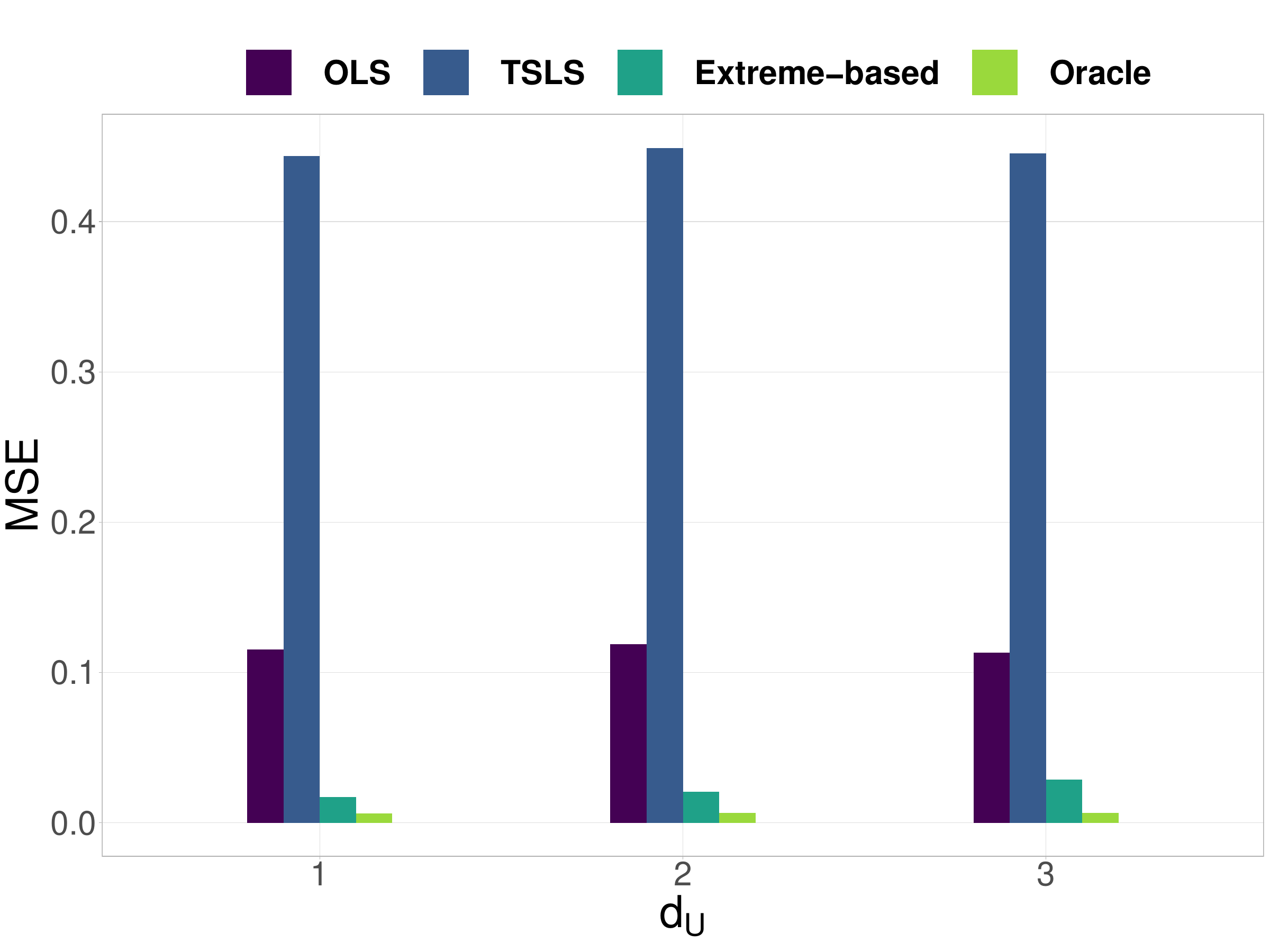}}
		\subfigure[Bias, $n = 5000$]{\includegraphics[scale = 0.18]{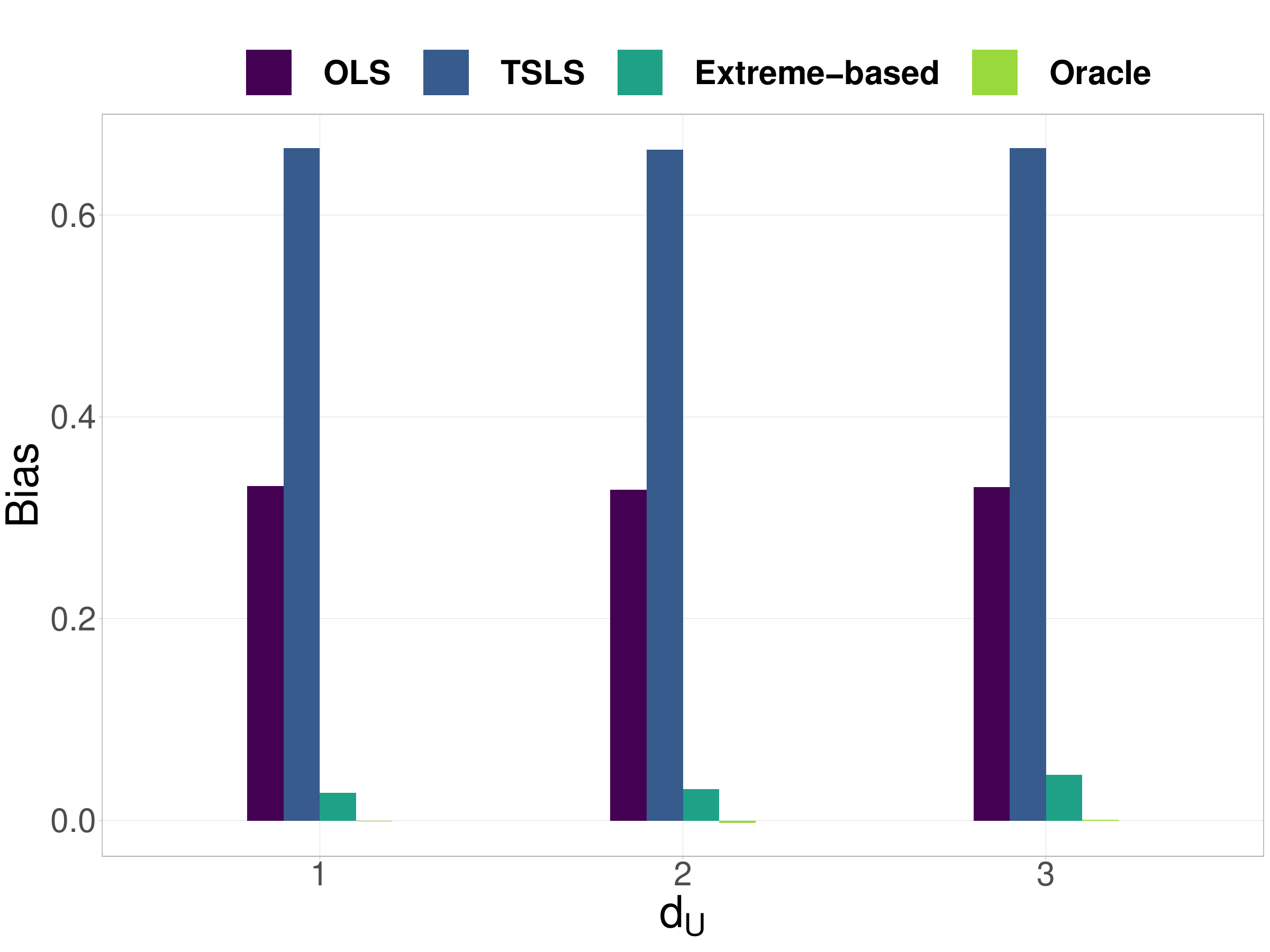}}
		\subfigure[MSE, $n = 5000$]{\includegraphics[scale = 0.18]{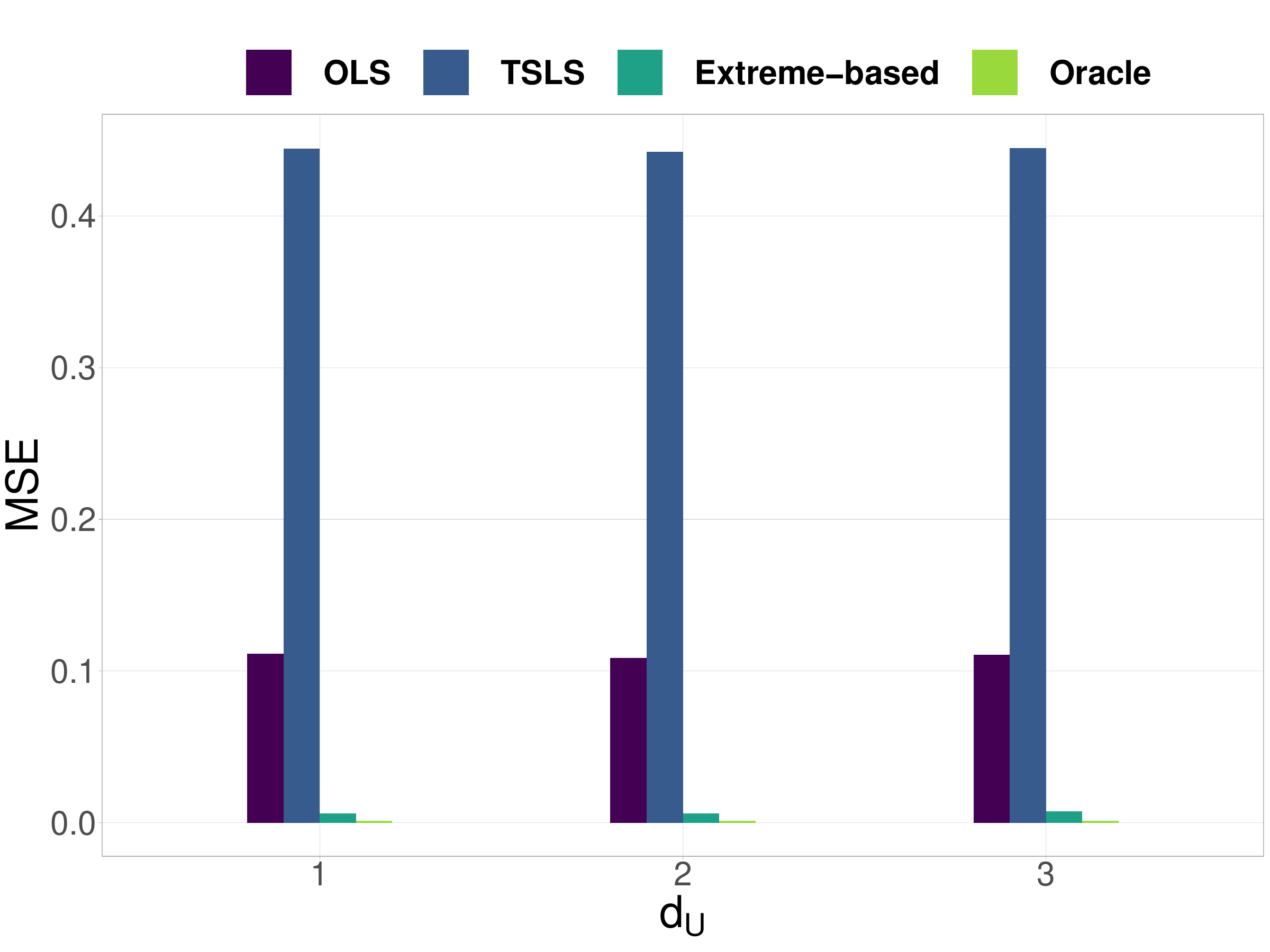}}
		\caption{Bias and MSE in the invalid IV problem with different $d_{U}$ and $n$.}\label{fig: estimation IV}
	\end{figure}

	Figure \ref{fig: estimation IV} presents the biases and MSEs of the OLS estimator, the TSLS estimator incorporating all candidate IVs, the proposed extreme-based estimator $\hat{\theta}_{n}$, and the oracle TSLS estimator using $Z_{1}$ as the IV and $Z_{2}, Z_{3}$ as measured confounder under different $d_{U}$ and $n$. 
	We adjust for $Z_{1}, Z_{2}$, and $Z_{3}$ in the implementation of the OLS estimator to account for their potential confounding effects. In the construction of $\hat{\theta}_{n}$, we first fit a linear regression between $X$ and $(Z_{1}, Z_{2}, Z_{3})$ and use the regression residual as the regressor in the subsequent quantile regression. This methodological adjustment aims to mitigate the influence of observed confounders, thereby making Assumption \ref{ass: balance} more plausible. The randomness of the regression coefficient between $X$ and $(Z_{1}, Z_{2}, Z_{3})$ is also taken into account in the bootstrap inference.
	
	Figure \ref{fig: estimation IV}  reveals that the OLS and TSLS estimators exhibit large biases and MSEs due to the unmeasured confounder and invalid IV problem, respectively. The extreme-based estimator $\hat{\theta}_{n}$ demonstrates substantially lower biases and MSEs than the OLS and TSLS estimators across different $d_{U}$ and $n$, highlighting its effectiveness and robustness in mitigating the effects of unmeasured confounding and invalid IVs. However, the bias and MSE of the extreme-based estimator are much larger than the oracle TSLS estimator due to its slow convergence rate. 
	
\begin{figure}[h]
		\centering
		\subfigure[Coverage rate, $n = 1000$]{\includegraphics[scale = 0.18]{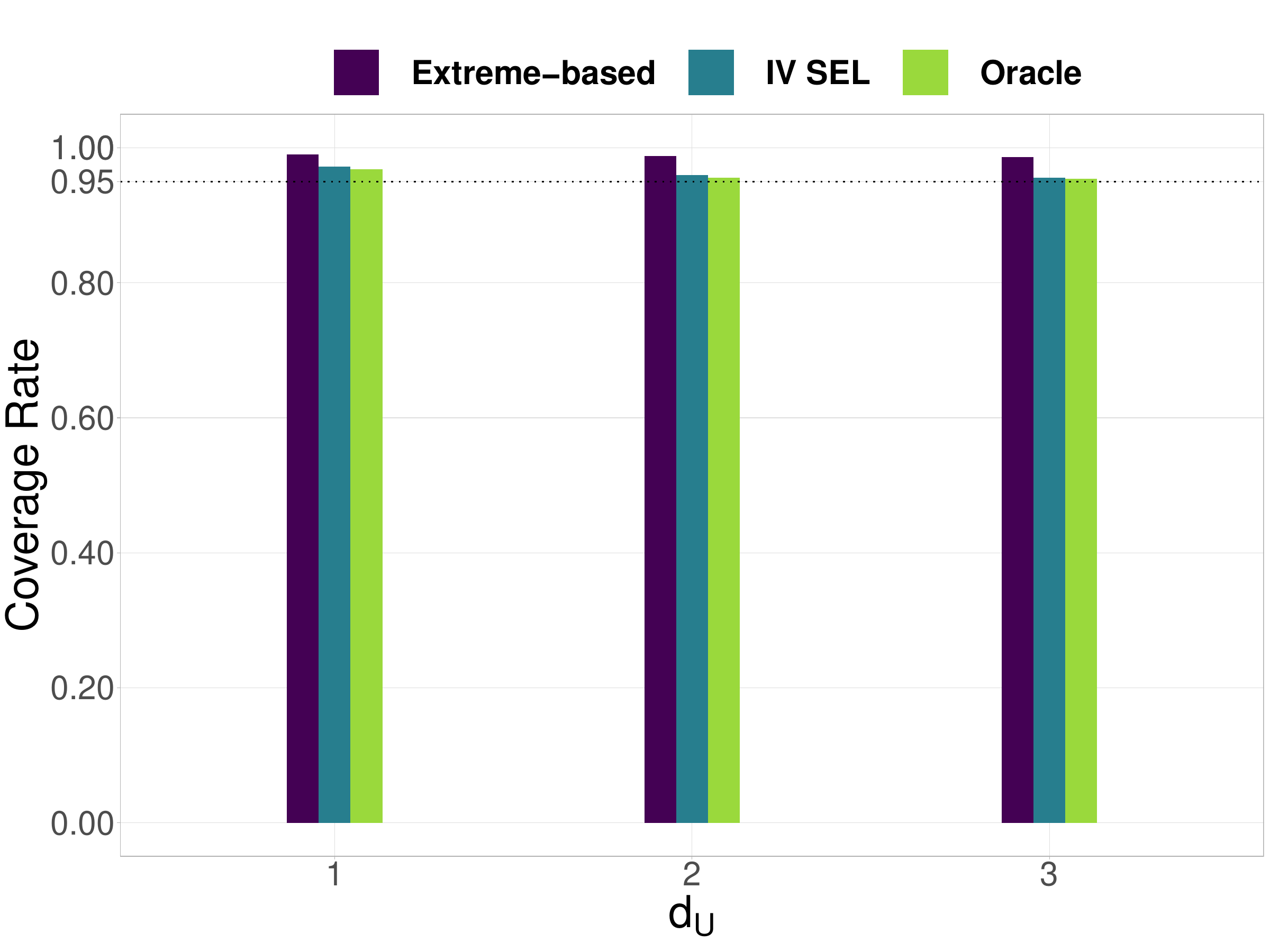}}
		\subfigure[Length, $n = 1000$]{\includegraphics[scale = 0.18]{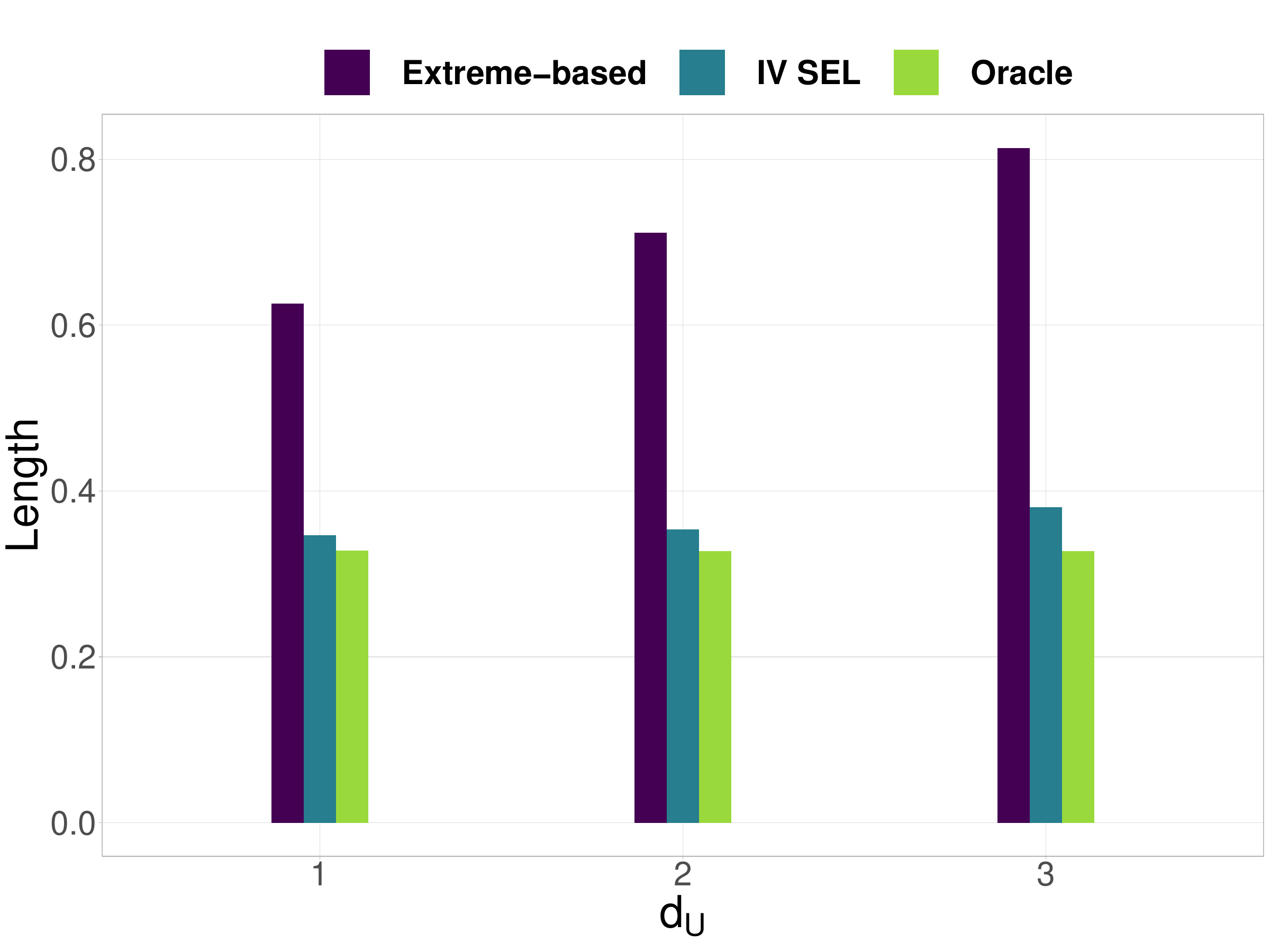}}
		\subfigure[Coverage rate, $n = 5000$]{\includegraphics[scale = 0.18]{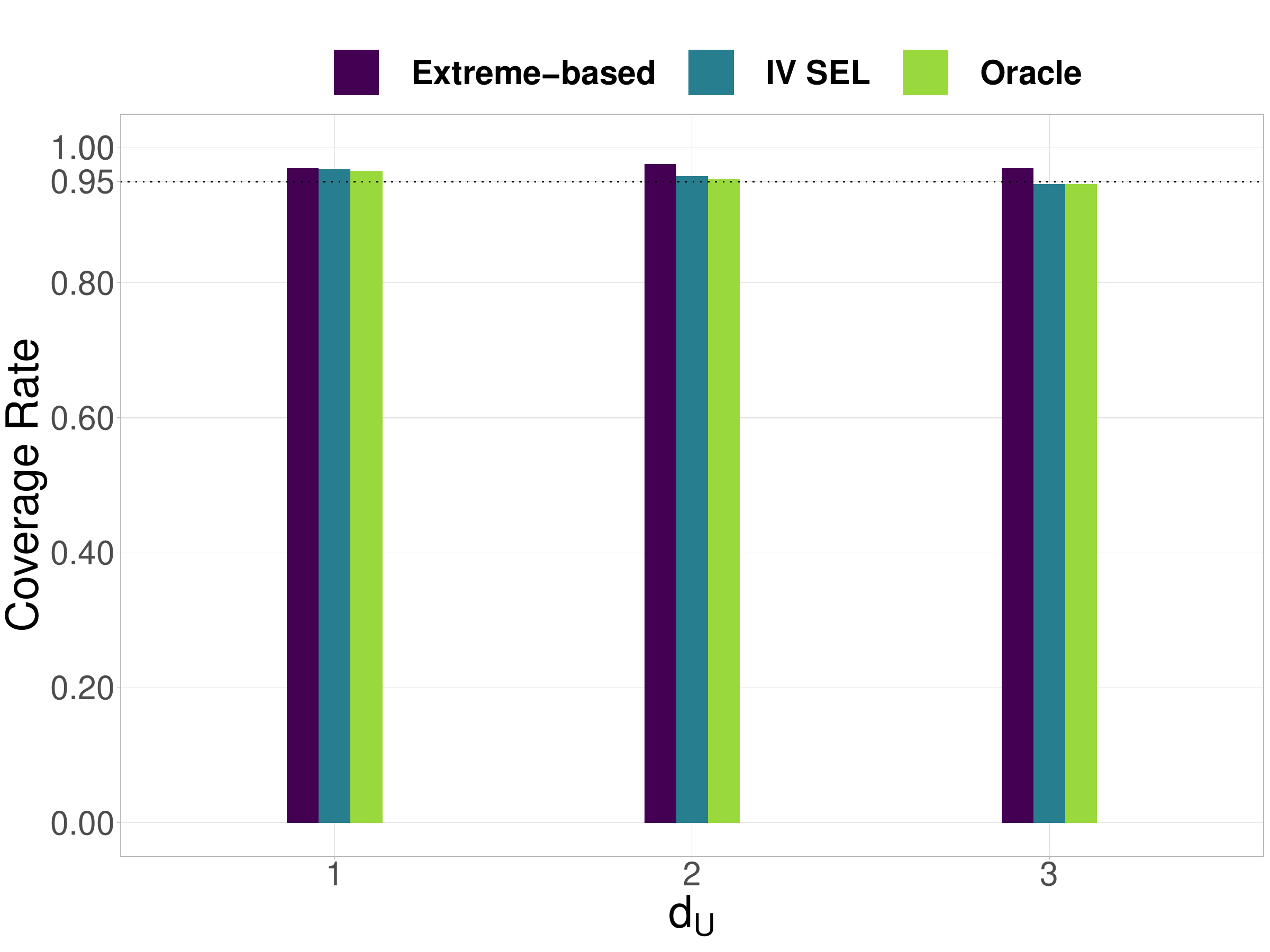}}
		\subfigure[Length, $n = 5000$]{\includegraphics[scale = 0.18]{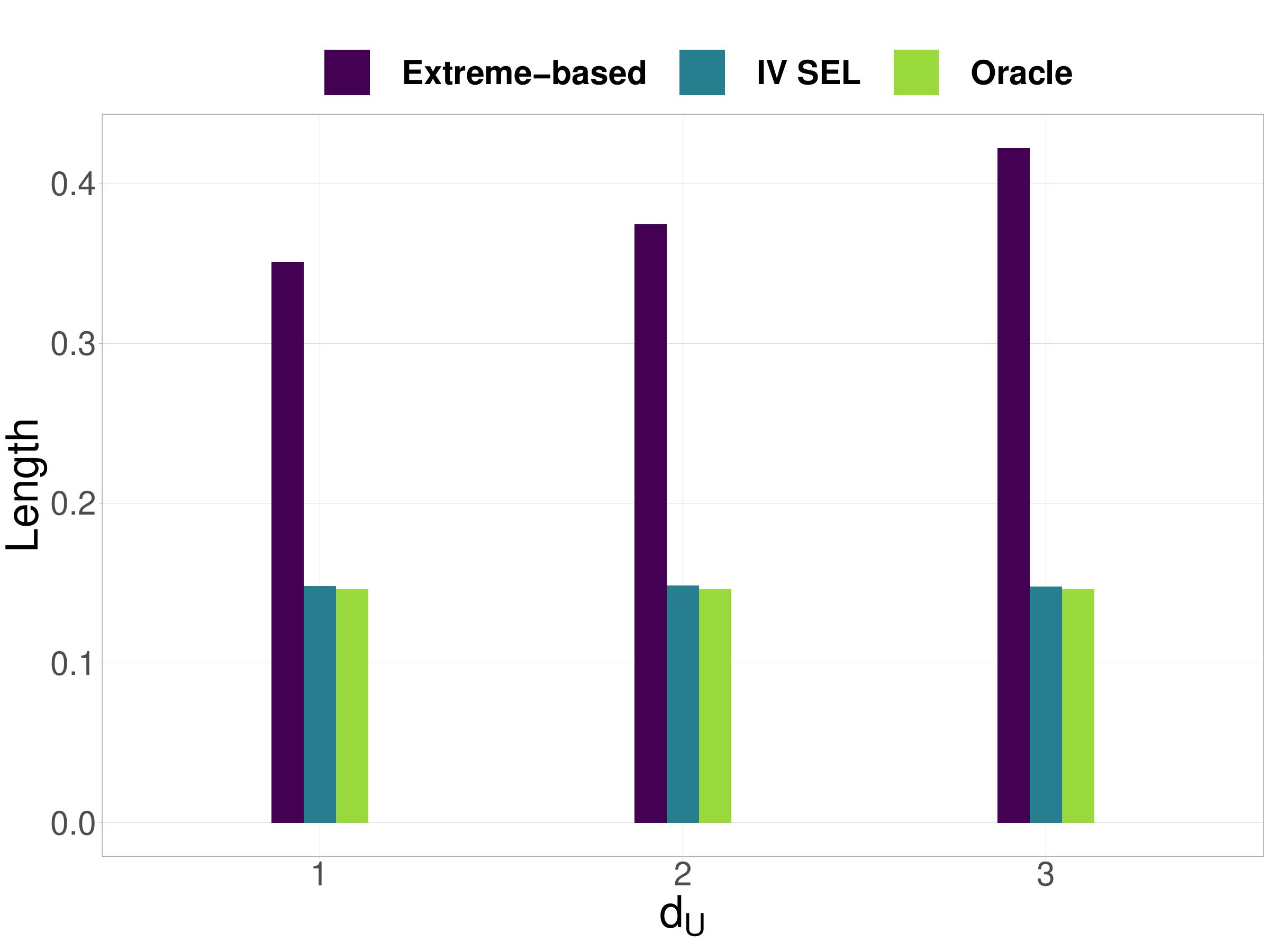}}
		\caption{Coverage rates and lengths of the $95\%$ confidence sets in the invalid IV problem with different $d_{U}$ and $n$.}\label{fig: CI IV}
	\end{figure}

	Figure \ref{fig: CI IV} presents the coverage rates and lengths of the $95\%$ confidence sets based on $\hat{\theta}_{n}$, as well as those constructed using selected IVs as detailed in Section \ref{app: inference invalid auxiliary} and the oracle TSLS. The candidate set of $\lambda$ for constructing \eqref{eq: IV CI} is $\{0.05, \dots, 0.95\}$ in the simulation. We do not include the confidence sets based on OLS and TSLS because their coverage rates are close to zero due to their large biases. 
	
	The coverage rates of all the confidence sets under comparison are larger than $0.95$ across all scenarios. By leveraging the IVs, the confidence sets based on selected IVs achieve shorter lengths than those based solely on extreme-based and have similar lengths as the confidence intervals based on the oracle TSLS when $n = 5000$.
\end{document}